\documentclass[aps,prd,onecolumn,superscriptaddress,nofootinbib,showpacs]{revtex4}

\usepackage{amsmath,bbm,latexsym,amssymb}

\usepackage{graphicx}

\usepackage{hyperref}
\hypersetup{colorlinks=true}

\def\be{\begin{equation}}
\def\ee{\end{equation}}
\def\ba{\begin{eqnarray}}
\def\ea{\end{eqnarray}}
\def\ra{\rightarrow}

\begin{document}
\preprint{CFTP/14-011}


\title{A reappraisal of the wrong-sign $hb\overline{b}$ coupling
and the study of $h \ra Z \gamma$}

\author{Duarte Fontes}\thanks{E-mail: duartefontes@tecnico.ulisboa.pt}
\affiliation{Centro de F\'{\i}sica Te\'{o}rica de Part\'{\i}culas (CFTP),
    Instituto Superior T\'{e}cnico, Universidade de Lisboa,
    1049-001 Lisboa, Portugal}
\author{J.~C.~Rom\~ao}\thanks{E-mail: jorge.romao@tecnico.ulisboa.pt}
\affiliation{Centro de F\'{\i}sica Te\'{o}rica de Part\'{\i}culas (CFTP),
    Instituto Superior T\'{e}cnico, Universidade de Lisboa,
    1049-001 Lisboa, Portugal}
\author{Jo\~{a}o P.~Silva}\thanks{E-mail: jpsilva@cftp.tecnico.ulisboa.pt}
\affiliation{Instituto Superior de Engenharia de Lisboa - ISEL,
	1959-007 Lisboa, Portugal}
\affiliation{Centro de F\'{\i}sica Te\'{o}rica de Part\'{\i}culas (CFTP),
    Instituto Superior T\'{e}cnico, Universidade de Lisboa,
    1049-001 Lisboa, Portugal}

\date{\today}

\begin{abstract}
It has been pointed out recently that current experiments
still allow for a two Higgs doublet model where the $h b \bar{b}$ coupling
($k_D m_b/v$) is negative;
a sign opposite to that of the Standard Model.
Due to the importance of delayed decoupling in
the $h H^+ H^-$ coupling,
$h \ra \gamma \gamma$ improved measurements will have
a strong impact on this issue.
For the same reason,
measurements or even bounds on
$h \ra Z \gamma$ are potentially interesting.
In this article,
we revisit this problem,
highlighting the crucial importance of $h \ra VV$,
which can be understood with simple arguments.
We show that the impacts on $k_D<0$ models
of both $h \ra b \bar{b}$ and $h \ra \tau^+ \tau^-$
are very sensitive to input values for the gluon fusion
production mechanism;
in contrast,
$h \ra \gamma \gamma$ and $h \ra Z \gamma$ are not.
We also inquire if the search for $h \ra Z \gamma$ and
its interplay with $h \ra \gamma \gamma$ will impact the
sign of the $h b \bar{b}$ coupling.
Finally,
we study these issues in the context of the Flipped two Higgs doublet model.
\end{abstract}

\pacs{12.60.Fr, 14.80.Ec, 14.80.-j}

\maketitle

\section{Introduction}

After the discovery of the Higgs particle by the
ATLAS \cite{atlas} and CMS  \cite{cms} experiments at LHC \cite{uptodate},
it became critically important to check how close its features
are to those in the Standard Model (SM).
Recently,
it has been emphasized by
Carmi \textit{et al.} \cite{Carmi:2012yp},
by Chiang and Yagyu \cite{Chiang:2013ixa}, by
Santos \cite{rui} and
by Ferreira \textit{et al.} \cite{FGHS}
that current data are consistent with
a lightest Higgs from a two Higgs doublet model
(2HDM) with a softly-broken $Z_2$ symmetry and CP conservation,
where the coupling of the bottom quark to the Higgs
($k_D\, m_b/v$) has a sign \textit{opposite}
to that in the SM.

Besides the SM gauge and fermion sector,
the model has two CP-even scalars,
$h$ and $H$,
one CP-odd scalar $A$,
and a conjugate pair of charged scalars $H^\pm$.
The scalar potential can be written in terms of
the usual vacuum expectation value (vev)
$v= 246$ GeV,
and seven parameters:
the four masses, $m_h$, $m_H > m_h$, $m_A$, and $m_{H^\pm}$;
two mixing angles, $\alpha$ and $\beta$,
and the (real) quadratic term breaking
$Z_2$, $m_{12}^2$.
With a suitable basis choice,
$\beta >0$ and $- \pi/2 \leq \alpha \leq \pi/2$.
Details about this model can be found,
for instances, in Refs.~\cite{HHG,PhysRep}.
We follow here the notation of the latter.

We concentrate on the Type II 2HDM,
where the fermion couplings with the lightest Higgs are
(multiplied by the mass of the appropriate fermion and divided by $v$)
\be
k_U = \frac{\cos{\alpha}}{\sin{\beta}},
\label{kU}
\ee
for the up-type quarks, and
\be
k_D = - \frac{\sin{\alpha}}{\cos{\beta}},
\label{kD}
\ee
for both the down-type quarks and the charged leptons.
In the SM limit, $k_U=k_D=1$.
Thus,
$\sin{\alpha}$ negative (positive) corresponds
to the (opposite of the) SM sign.
The couplings of $h$ to vector boson pairs are
\be
g_{hVV} = k_V g_{hVV}^{\textrm{SM}} =
\sin{(\beta-\alpha)}\  g_{hVV}^{\textrm{SM}},
\label{ghVV}
\ee
where $VV = ZZ, W^+ W^-$,
and the coupling to a pair of charged Higgs bosons may be written
as \cite{Posch}
\be
g_{h H^+ H^{-}}
= - \frac{2\, m_{H^\pm}^2}{v}
\left(
g_1 + g_2 + g_3
\right),
\label{ghH+H-}
\ee
where
\ba
g_1
&=&
\sin{(\beta-\alpha)}
\left(
1 - \frac{m_h^2}{2 m_{H^\pm}^2}
\right),
\nonumber\\
g_2
&=&
\frac{\cos{(\beta+\alpha)}}{\sin{(2 \beta)}}
\frac{m_h^2}{m_{H^\pm}^2},
\nonumber\\
g_3
&=&
- \frac{2 \cos{(\beta+\alpha)}}{\sin^2{(2 \beta)}}
\frac{m_{12}^2}{m_{H^\pm}^2}.
\ea
We have checked,
with the help of
\texttt{FeynRules}~\cite{FeynRules},
that this expression is correct.

Before $H^\pm$ are detected directly, their effect might be detected
indirectly through loop contributions involving $g_{h H^+ H^{-}}$,
especially in decays of $h$ which are already loop decays in the SM,
such as $h \ra \gamma \gamma$ and $h \ra Z \gamma$.  This is possible
if $m_{H^\pm} \sim v$, because there will be a light particle in the
loop.  This is also possible for $m_{H^\pm} \gg v$, when the
$H^\pm$ loop contribution approaches a constant
\cite{Arhrib:2003ph,Bhattacharyya:2013rya}.  However, making
$m_{H^\pm}$ too large will require quartic couplings in violation of
the unitarity bounds.  This still leaves a rather wide range of
$H^\pm$ masses where the charged Higgs contributions to $h \ra \gamma
\gamma$ and $h \ra Z \gamma$ could be detected.  In Ref.~\cite{FGHS},
it is shown that such non-decoupling is unavoidable in $h \ra \gamma
\gamma$, if the wrong-sign $hb\overline{b}$ ($k_D < 0$) case is to
conform to all current data.  We have checked that such non-decoupling
will also have an impact on $h \ra Z \gamma$.

Our article is organized as follows.  In Section~\ref{sec:fit}, we
discuss our fit procedure.  There are differences with respect to
Ref.~\cite{FGHS}, most notably in the production rates, as shown in
section~\ref{subsec:differ}.  In section~\ref{subsec:VV}, we point out
the crucial importance of the $VV$ channel by itself, which can be
understood with quite simple arguments.  It turns out that, once $VV$
is constrained, $b \bar{b}$ and $\tau^+ \tau^{-}$ are rather sensitive
to the production rates, while $\gamma \gamma$ and $Z \gamma$ are not.
This feature is explained in detail in
section~\ref{subsec:production}.  Section~\ref{sec:14} includes our
predictions for the next LHC run, which will occur at 14 TeV (not 8
TeV).
We show that, before applying the constraints on $VV$,
$Z \gamma$ can be above the SM value by a factor of two.
If such values were to be measured, we would exclude the SM.
However, $Z \gamma$ can also take the SM value, and it cannot be used
to exclude $k_D < 0$.  In section~\ref{sec:flipped}, we analyze the
Flipped 2HDM, where the coupling to the charged leptons goes like
$k_U$ in Eq.~\eqref{kU} -- not like $k_D$ in Eq.~\eqref{kD}.  We draw
our conclusions in Section~\ref{sec:conclusions}.

\section{\label{sec:fit} Fit procedure and some results}

The scalar particle found at the LHC has been seen in
the $\gamma \gamma$, $Z Z^\ast$, $W W^\ast$,
and $\tau^+ \tau^-$ final states,
with errors of order $20\%$.
The $b \bar{b}$ final state is only seen
(at LHC and the Tevatron)
in the associated $Vh$ production mechanism,
with errors of order $50\%$
\cite{cms:bb, Tuchming:2014fza}.
Searches have also been performed for the
$Z \gamma$ final state \cite{atlas:Zph, cms:Zph},
with upper bounds around ten times the SM expectation
at the $95\%$ confidence level.
Current LHC results can be found in Ref.~\cite{uptodate}.

These results for the
$pp \ra h \ra f$ rates (where $f$ is some final state)
are usually presented in the form of
ratios of observed rates to SM expectations.
This is what we use to constrain the ratios between
the 2HDM and SM rates
\be
\mu_f
=
R_P\, R_D\, R_{TW},
\label{mus}
\ee
where the sub-indices $P$, $D$, and $TW$
stand for ``production'', ``decay'', and ``total width'',
respectively.
Here,
\ba
R_P
&=&
\frac{\sigma^\textrm{2HDM}(pp \ra h)}{\sigma^\textrm{SM}(pp \ra h)},
\label{RP}
\nonumber\\*[3mm]
R_D
&=&
\frac{\Gamma^\textrm{2HDM}[h \ra f]}{\Gamma^\textrm{SM}[h \ra f]},
\label{RD}
\nonumber\\*[3mm]
R_{TW}
&=&
\frac{\Gamma^\textrm{SM}[h \ra \textrm{all}]}{
\Gamma^\textrm{2HDM}[h \ra \textrm{all}]},
\label{RTW}
\ea
where $\sigma$ is the Higgs production mechanism,
$\Gamma[h \ra f]$ the decay width into the final state $f$,
and $\Gamma[h \ra \textrm{all}]$ is the total Higgs decay width.

We follow the strategy of Ref.~\cite{FGHS},
and assume that all observed decays have been measured at the SM rates,
with the same error $20\%$.
For the most part,
we keep $b \bar{b}$ out of the mix,
because: it has larger errors;
it is only measured in the $Vh$ production channel;
and, as we will show,
it is not needed in Type II models,
were $\tau^+ \tau^-$ has the same effect
(which, moreover, is not very large).
We will only assume that all production mechanisms
are involved in $b \bar{b}$ and that its errors are of order $20\%$
when we wish to compare with Ref.~\cite{FGHS},
explaining the differences in production.

We have performed extensive simulations of the type II 2HDM,
with the usual strategy. We set $m_h = 125 $ GeV,
generate random points for $-\pi/2 \leq \alpha \leq \pi/2$,
$ 1 \leq \tan{\beta} \leq 30$,
$ 90\ \textrm{GeV} \leq m_A \leq 900\ \textrm{GeV}$,
$ 125\ \textrm{GeV} \leq m_H \leq 900\ \textrm{GeV}$,
$ - (900\ \textrm{GeV})^2 \leq m_{12}^2 \leq (900\ \textrm{GeV})^2$,
and $340\ \textrm{GeV} \leq m_{H^\pm} \leq 900\ \textrm{GeV}$.
These coincide with the ranges in Ref.~\cite{FGHS},
where $\tan{\beta}$ and $m_{H^\pm}$ were chosen
to conform with $B$ Physics and $Z \ra b \bar{b}$ data.

For each point,
we derive the parameters of the scalar potential,
and we keep only those points which provide a bounded from below
solution \cite{Deshpande:1977rw},
respecting perturbative unitarity
\cite{Kanemura:1993hm, Akeroyd:2000wc,Ginzburg:2003fe},
and the constraints from the oblique radiative
parameters $S, T, U$ \cite{Grimus:2008nb, Baak:2012kk}.
At the end of this procedure,
we have a set of possible 2HDM parameters,
henceforth denoted simply by SET.

Next, we generate the rates for all channels,
including all production mechanisms:
$gg \ra h$  (gluon fusion) at NNLO from HIGLU \cite{Spira:1995mt},
$b \bar{b} \ra h$ at NNLO from SusHi \cite{Harlander:2012pb},
$Vh$ associated production,
$t \bar{t} h$,
and $VV \ra h$ (vector boson fusion) \cite{LHCCrossSections}.
In the SM, the production cross section is dominated by the gluon
fusion process with internal top quark.  Generically speaking, this
also holds in the Type II 2HDM, but, given Eq.~\eqref{kD}, the
contribution from the gluon fusion process with internal bottom quark
becomes more important as $\tan{\beta}$ increases.

\subsection{\label{subsec:differ}Comparing with previous results}

\begin{figure}[htb]
\centering
\begin{tabular}{cc}
\includegraphics[width=0.49\textwidth]{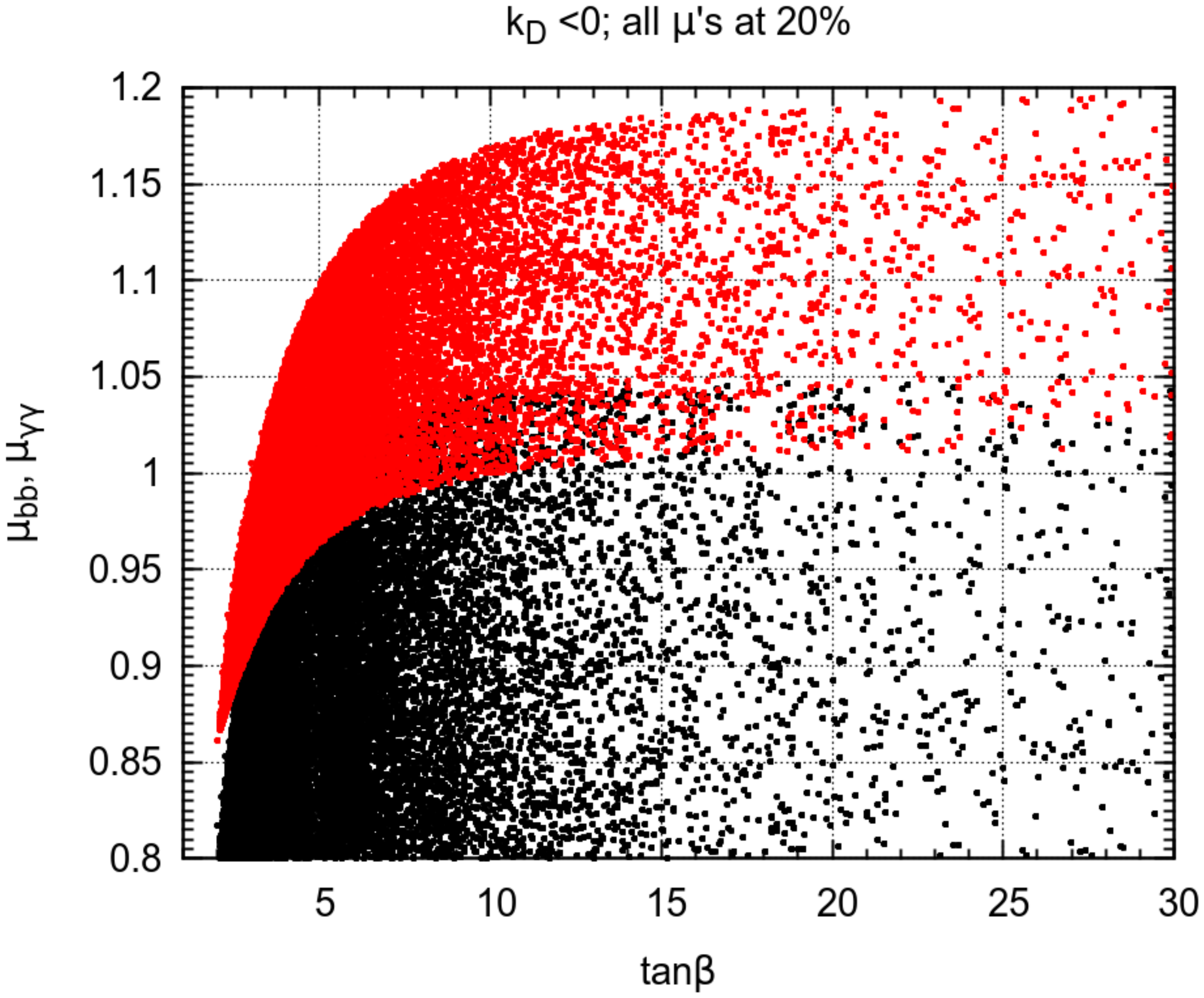} &
\includegraphics[width=0.49\textwidth]{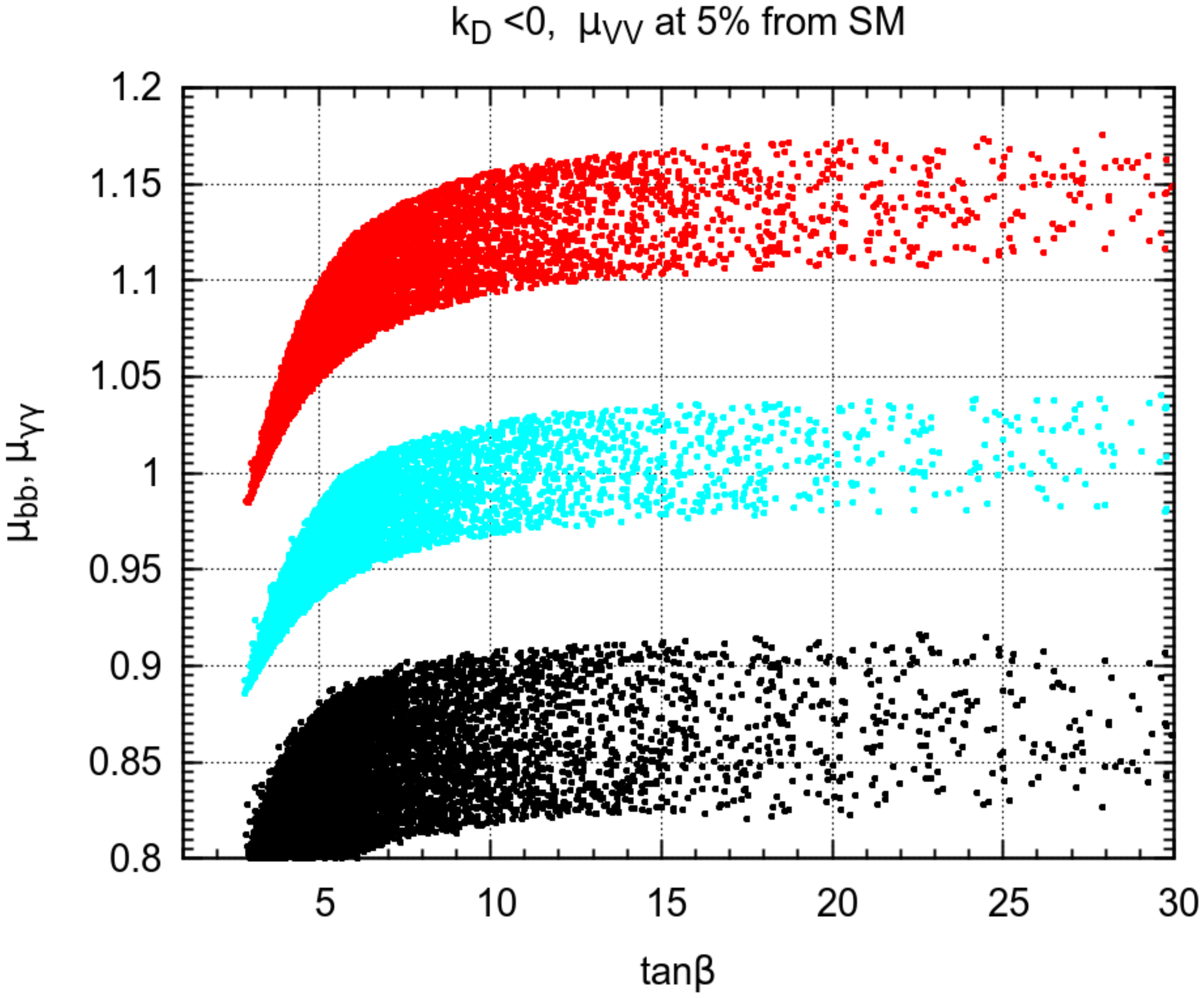}
\end{tabular}
\caption{Left panel: Assuming that all $\mu_f$ are within 20\% of the SM
  prediction we plot $\mu_{b \bar{b}}$ (red/dark-gray) and
  $\mu_{\gamma\gamma}$ (black). Right panel: Assuming now that
  $\mu_{VV}$ are within 5\% of the SM prediction we plot the same
  quantities. For comparison we also plot $\mu_{b \bar{b}}$
  (cyan/light-gray) for the assumed production of Ref.~\cite{FGHS}.}
\label{fig:5L}
\end{figure}

We start by requiring that all
points in the SET obey $k_D<0$ and $0.8 \leq \mu_f \leq 1.2$ for the $VV$, $\tau^+
\tau^-$, $b \bar{b}$, and $\gamma \gamma$ at 8 GeV.  The surviving
points are plotted as a function of $\tan{\beta}$ in the left panel of
Fig.~\ref{fig:5L}, where we show the possible values of
$\mu_{\gamma\gamma}$ (in black) and $\mu_{b \bar{b}}$ (in red/dark-gray).
We notice that, although  $\mu_{b \bar{b}}$ is in general larger than
$\mu_{\gamma\gamma}$, the two regions overlap and, at the level of
deviations of 20\% from the SM, both are compatible with the SM value of one.
We now assume that $h\to VV$ is
measured within 5\% of the SM:
$0.95 \leq \mu_{VV} \leq 1.05$.
The result is plotted in the right panel of Fig~\ref{fig:5L}.
We found that $\mu_{\gamma\gamma}$ agrees, within
errors, with that shown in Fig.~5-Left of Ref.~\cite{FGHS},
while our result for $\mu_{b \bar{b}}$ is well above theirs,
which we also show in Fig.~\ref{fig:5L} (in
cyan/light-gray).
This is puzzling, since we can reproduce their remaining
plots.

After comparing notes with R.~Santos from Ref.~\cite{FGHS}, we found
that the difference originates in the gluon fusion production rates,
because we are using values from an more recent version of
HIGLU \cite{Spira:1995mt}, and, eventually,
different PDF's and energy scales.  For example, they quote
\be
\frac{\sigma(gg \ra h)^\textrm{2HDM}_\textrm{NNLO}}{
\sigma(gg \ra h_\textrm{SM})_\textrm{NNLO}}
=
1.06
\hspace{3ex}
[\sin{(\beta+\alpha)=1}],
\ee
while we,
using the latest version (4.0) of
HIGLU \cite{Spira:1995mt},
obtain
\be
\frac{\sigma(gg \ra h)^\textrm{2HDM}_\textrm{NNLO}}{
\sigma(gg \ra h_\textrm{SM})_\textrm{NNLO}}
=
1.126
\hspace{3ex}
[\sin{(\beta+\alpha)=1}].
\ee
This apparently explains why our $\mu_{b \bar{b}}$ result (in red/dark-gray) lies
above the one which we obtain (in cyan/light-gray) with the
assumed production rates used in Ref.~\cite{FGHS}.

But, this raises another puzzle. If the only difference lies in the
production rates, why do our results for $\mu_{\gamma\gamma}$
agree with those in Ref.~\cite{FGHS}?  This is what we turn to in
section~\ref{subsec:production}.

\subsection{\label{subsec:VV} The crucial importance of $h \rightarrow VV$
and trigonometry}

In the previous section, we required that all points obey $0.8 \leq
\mu_f \leq 1.2$ for all final states $VV$, $\tau^+ \tau^-$, $b
\bar{b}$, and $\gamma \gamma$, simultaneously.  The problem with this
procedure is that one misses out on the crucial importance that
$\mu_{VV}$ has on its own.

In this section, we only assume that $0.8 \leq \mu_{VV} \leq 1.2$, and
we will make the cavalier assumption that the production is due
exclusively to the gluon fusion with intermediate top, while the decay
is due exclusively to the decay $h \ra b \bar{b}$ \cite{degenerate}.
Under these assumptions,
\be
\mu_{VV}
\approx
\frac{k_U^2}{k_D^2} \sin^2{(\beta - \alpha)}.
\label{muVV_approx}
\ee
We now perform a simple trigonometric exercise.  We vary $\alpha$
between $-\pi/2$ and $\pi/2$, $\tan{\beta}$ between $1$ and $30$, and
we only keep those regions where $0.8 \leq \mu_{VV} \leq 1.2$, with
the approximation in Eq.~\eqref{muVV_approx}.

\begin{figure}[htb]
\centering
\begin{tabular}{cc}
\includegraphics[width=0.49\textwidth]{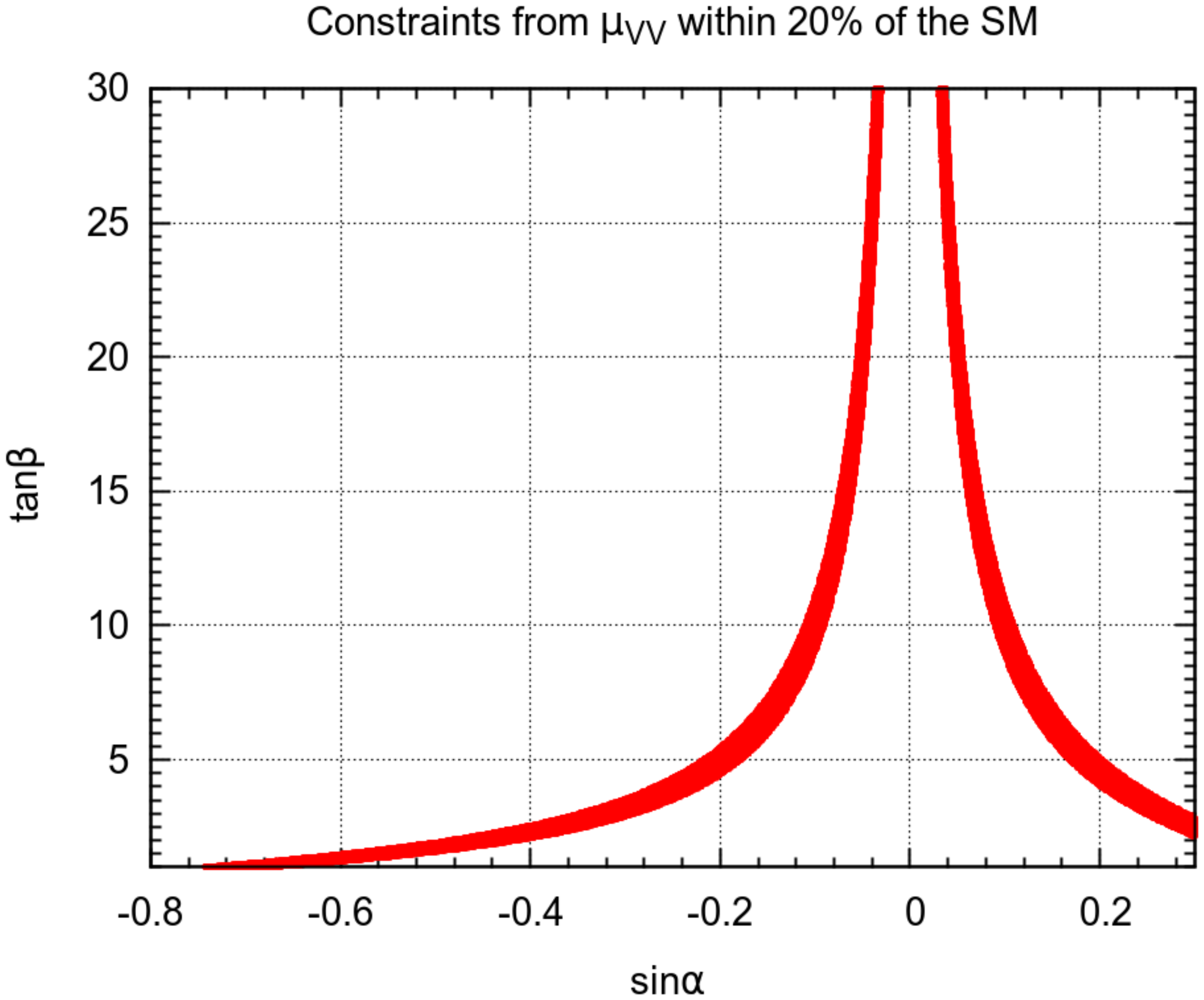}&
 \includegraphics[width=0.49\textwidth]{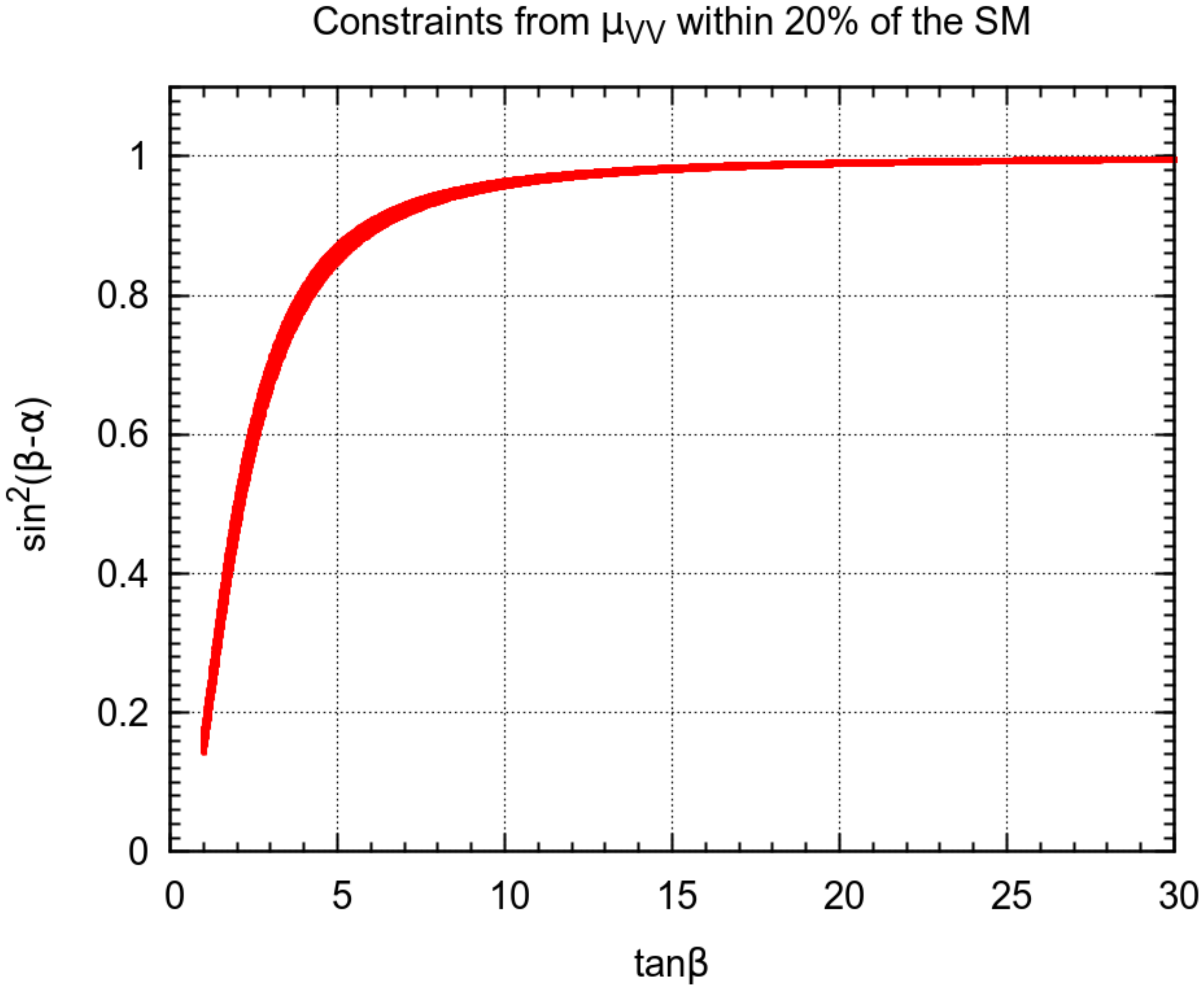}
\end{tabular}
\caption{Left panel: Plot of $\tan\beta$ as a function of $\sin\alpha$
  for all the points that obey Eq.~\eqref{muVV_approx} with $0.8 \leq
  \mu_{VV} \leq 1.2$. Right panel: Plot of $\sin^2(\beta-\alpha)$ as a
  function of $\tan\beta$ for the points that obey
  Eq.~\eqref{muVV_approx} with $0.8 \leq \mu_{VV} \leq 1.2$ and have $k_D<0$.}
\label{fig:2L}
\end{figure}

In Fig.~\ref{fig:2L}, we show the remaining points in the
$\sin{\alpha} - \tan{\beta}$ plane.
This matches remarkably well the
Fig.~2-Left from Ref.~\cite{FGHS}.
That is, a simple back of the envelope calculation has most of the
Physics.  The left branch of the left panel of Fig.~\ref{fig:2L}
corresponds to the SM sign ($k_D > 0$), and it lies very close to the
curve $\sin{(\beta-\alpha)}=1$.  The right branch of the same figure
corresponds to the wrong sign ($k_D < 0$), and lies very close to the
curve $\sin{(\beta+\alpha)}=1$ \cite{rui}.

\begin{figure}[htb]
\centering
\includegraphics[width=0.60\textwidth]{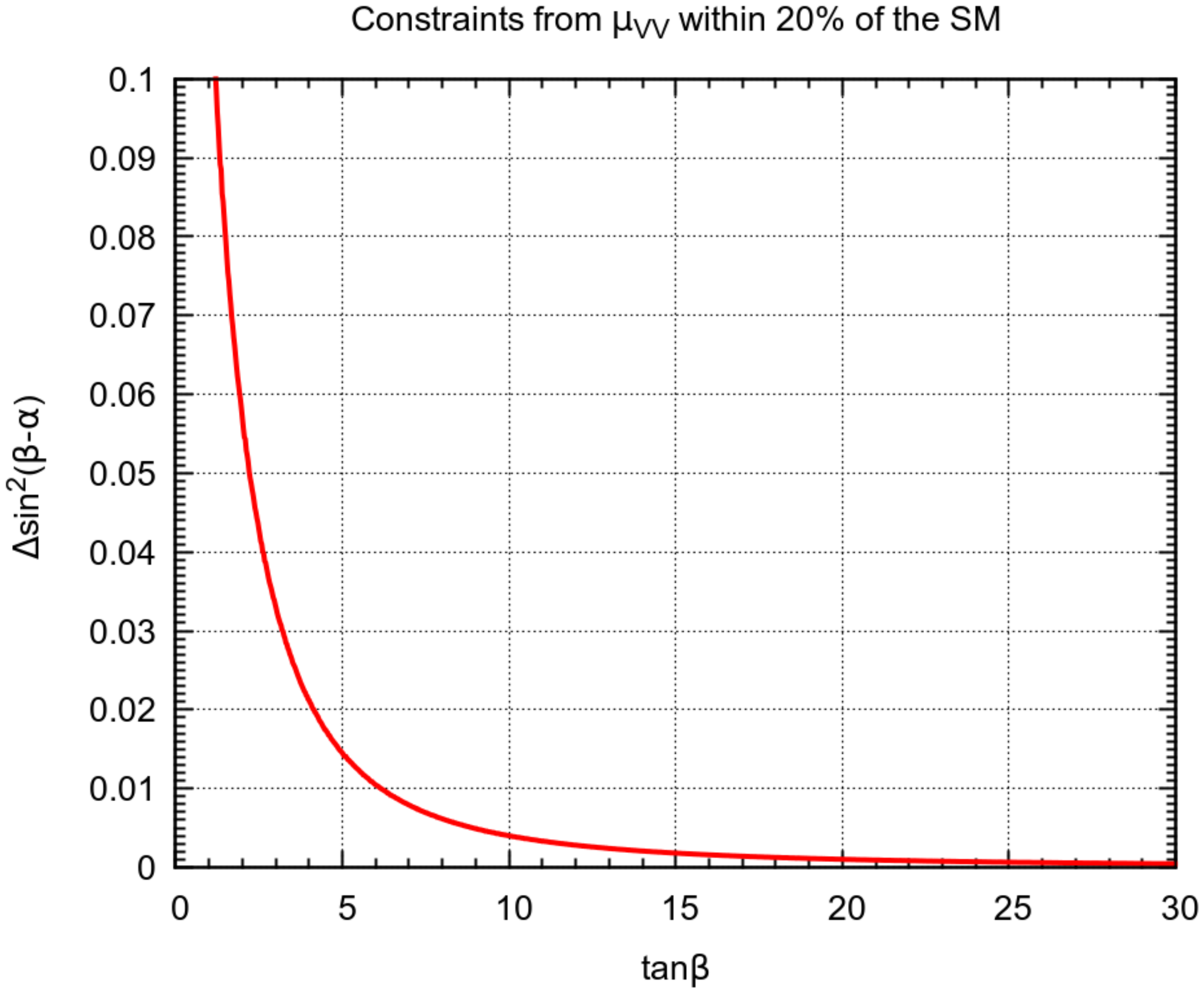}
\caption{Fractional variation of $\sin^2{(\beta-\alpha)}$  as a
  function of $\tan\beta$ for all points with $k_D<0$ that obey
  Eq.~\eqref{muVV_approx} with $0.8 \leq  \mu_{VV} \leq 1.2$.}
\label{fig:crazy2}
\end{figure}

Under the same assumptions, we can draw $\sin^2{(\beta-\alpha)}$ as a
function of $\tan{\beta}$, as seen in the right panel of
Fig.~\ref{fig:2L}, keeping only $\sin{\alpha} >0$ ($k_D <0$) points.
Notice that $\sin^2{(\beta-\alpha)}$ becomes almost univocally defined
in terms of $\tan{\beta}$.  Indeed, fixing $\tan{\beta}$, and defining
the fractional variation of $\sin^2{(\beta-\alpha)}$ around its
average value by
\be
\Delta =
\frac{\sin^2{(\beta-\alpha)}_\textrm{max} -
\sin^2{(\beta-\alpha)}_\textrm{min}}{
\sin^2{(\beta-\alpha)}_\textrm{max} +
\sin^2{(\beta-\alpha)}_\textrm{min}},
\ee
we obtain the results in Fig.~\ref{fig:crazy2}.
For small $\tan{\beta}$, $\sin^2{(\beta-\alpha)}$ is determined to
better than $10\%$, when $\mu_{VV}$ is fixed only to $20\%$ accuracy.
Although it might seem from Eq.~\eqref{muVV_approx} that it should be
roughly the same, it turns out that the inclusion in
Eq.~\eqref{muVV_approx} of $k_U$ and $k_D$ from
Eqs.~\eqref{kU}-\eqref{kD} helps in reducing the error.  But things
get even more accurate as $\tan{\beta}$ increases.  For example, for
$\tan{\beta}=10$, $\sin^2{(\beta-\alpha)}$ differs very little from
unity, and it is even more precisely defined around its average value;
an accuracy better than $0.5\%$, coming from a $\mu_{VV}$ fixed only
to $20\%$ accuracy.
\begin{figure}[hb]
\centering
\begin{tabular}{cc}
\includegraphics[width=0.48\textwidth]{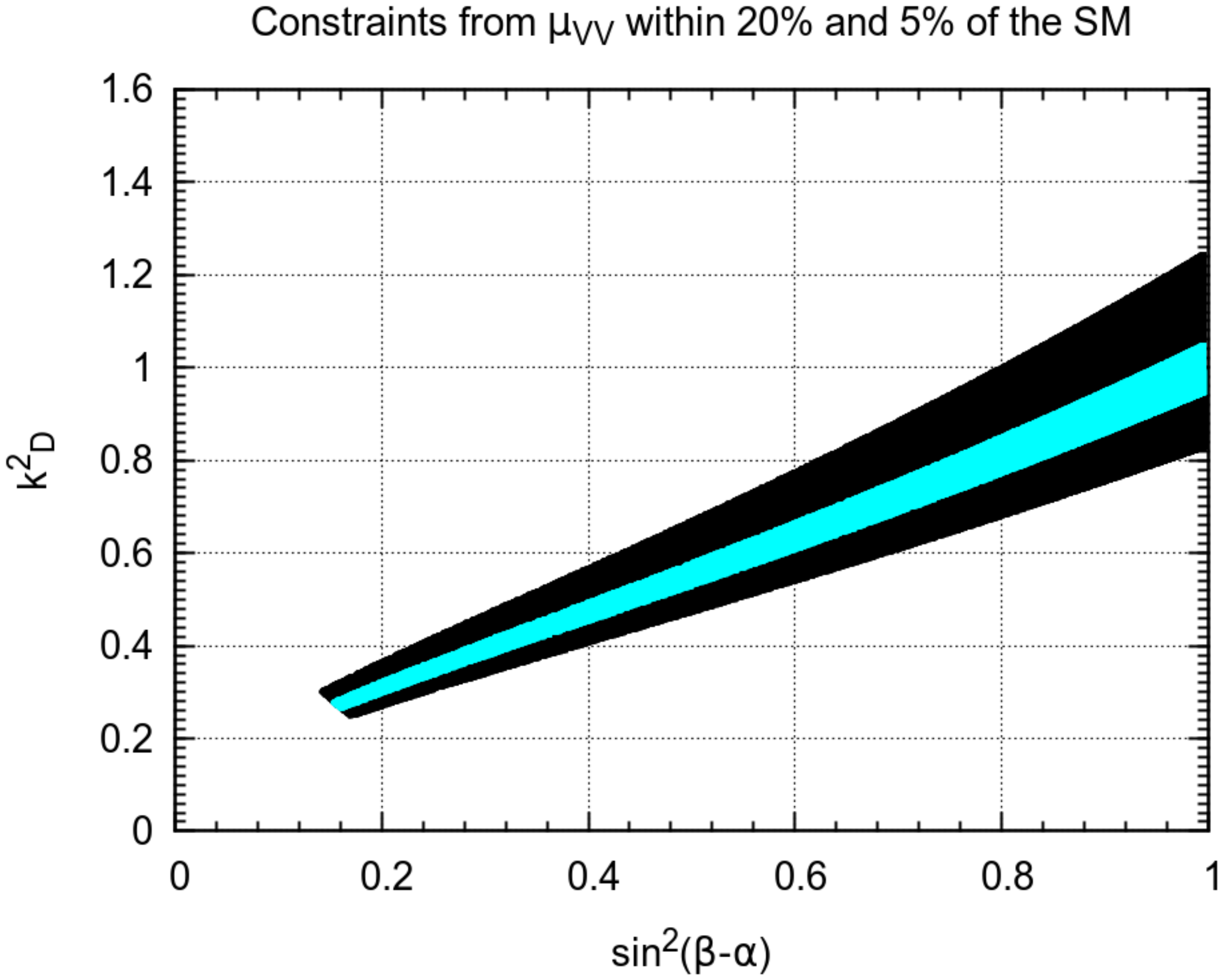}&
\includegraphics[width=0.48\textwidth]{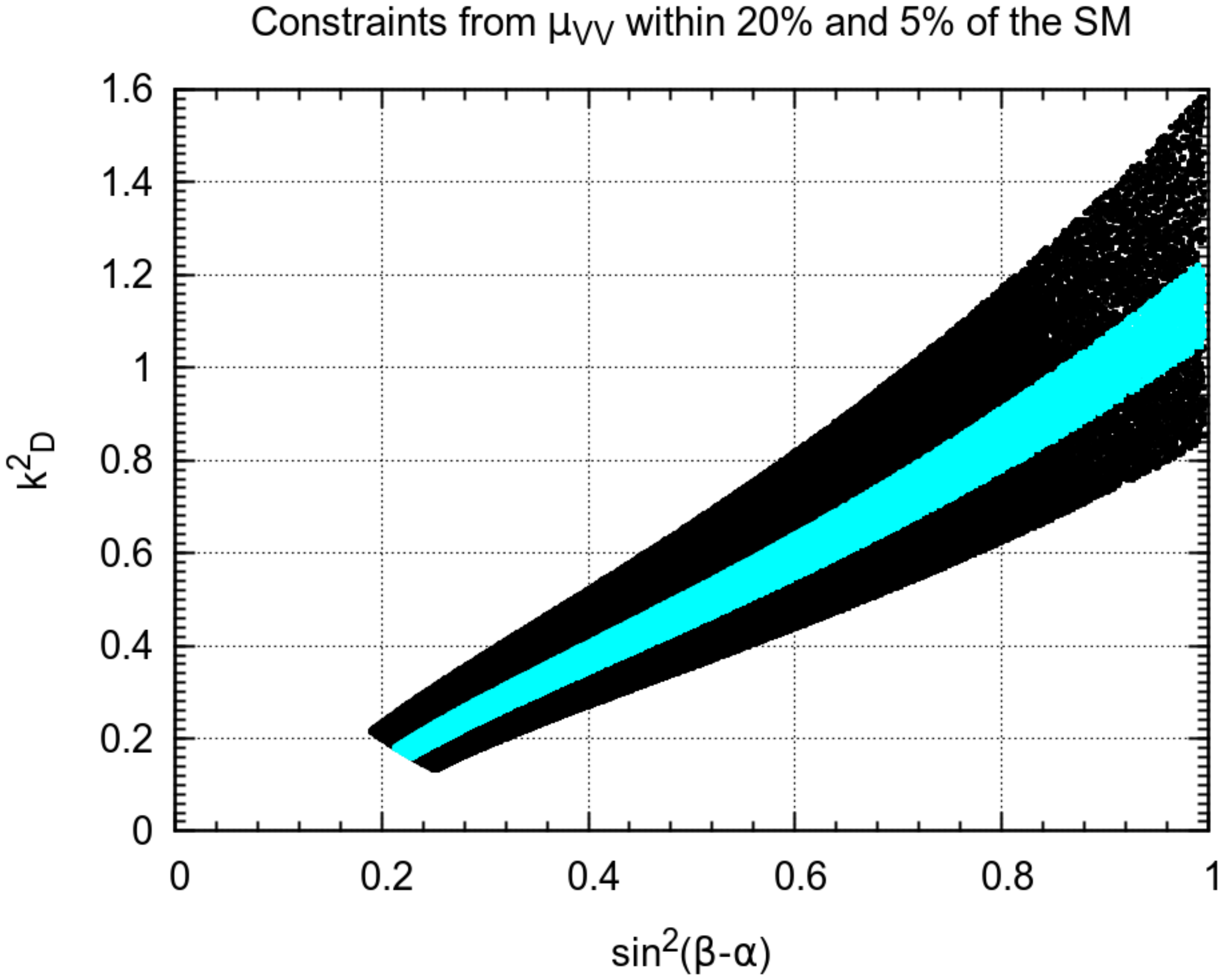}
\end{tabular}
\caption{Left panel: $k^2_D$ as a function of  $\sin^2{(\beta-\alpha)}$ for all
  points with $k_D<0$ that obey
  Eq.~\eqref{muVV_approx} with $0.8 \leq  \mu_{VV} \leq 1.2$ (black)
  or with $0.95 \leq  \mu_{VV} \leq 1.05$ (cyan/light-gray). Right
  panel: The same but for generated data obeying the model constraints.}
\label{fig:crazy3}
\end{figure}

Finally, in the left panel of Fig.~\ref{fig:crazy3} we show $k_D^2$ as
a function of $\sin^2{(\beta-\alpha)}$, under the same assumptions
(black). For comparison, we show how this relation becomes more
constrained if we require $0.95 \leq \mu_{VV} \leq 1.05$ (cyan/light-gray).
To emphasize that the trigonometric relations which result from
$\mu_{VV}$ in Eq.~\eqref{muVV_approx} explain most of the results, we
show in the right panel of Fig.~\ref{fig:crazy3} the same plot but now
with points generated obeying all the model constraints and without
the simplifying assumptions that led to Eq.~\eqref{muVV_approx}.
These simple considerations will turn out to be very important in the
next section.

\subsection{\label{subsec:production} How production affects the rates}

In the previous section, we have made a drastic approximation, which
reduced the analysis to a simple trigonometric issue in $\alpha$ and
$\beta$, with no dependence on other 2HDM parameters.  Now we resume
the SET found by scanning all the 2HDM parameter space and imposing
theoretical constraints, as defined at the beginning of
section~\ref{sec:fit}; we then use all production mechanisms.

\begin{figure}[h]
\centering
\includegraphics[width=0.6\textwidth]{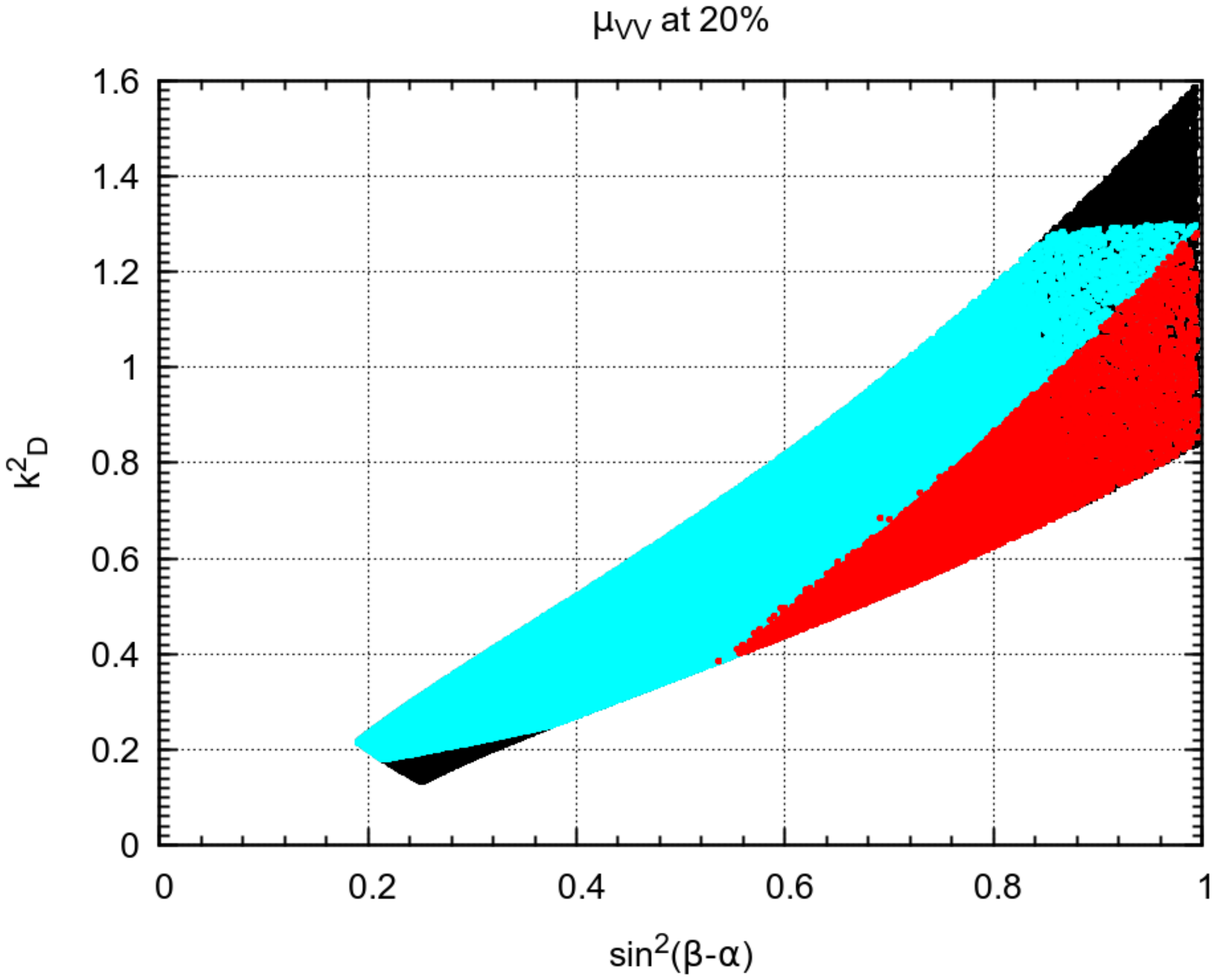}
\caption{Allowed region for $k^2_D$ as a function of  $\sin^2{(\beta-\alpha)}$ for all
  points with $k_D<0$ that obey $0.8 \leq  \mu_{VV} \leq 1.2$
  (black). The region in cyan (light-gray)
  is obtained by imposing in addition that
  $0.8 \leq  \mu_{\tau\tau} \leq 1.2$, while in the region in red (dark-gray) we further impose  $0.8 \leq  \mu_{\gamma\gamma} \leq 1.2$. }
\label{fig:KD2sin2bma}
\end{figure}

In Fig.~\ref{fig:KD2sin2bma}, we show our 8 TeV results for $k_D^2$ as
a function of $\sin^2{(\beta-\alpha)}$.  In black, we see
the points generated from the SET, constrained exclusively by $0.8
\leq \mu_{VV} \leq 1.2$.
This coincides with the black region in the right panel of
Fig.~\ref{fig:crazy3}, and should be compared with the left panel of
Fig.~\ref{fig:crazy3}. As already mentioned, the similarity is uncanny.
Simple trigonometry really does have a very strong impact on the
results, particularly in the values of $\sin^2{(\beta-\alpha)}$; its
ranges are practically the same in the two figures.  The value for
$k_D^2$ for low $\sin^2{(\beta-\alpha)}$ (which, as we see from the
right panel of Fig.~\ref{fig:2L}, occurs for low $\tan{\beta}$), is
also rather similar.  There are, of course, minor quantitative
differences: some due to the fact that the SET already has some
constraints on the model parameters, due to the imposition of the
bounded from below, perturbativity and $S$, $T$, $U$ conditions; some
due to the details of the production mechanism.  The most important
difference occurs for $\sin^2{(\beta-\alpha)} \sim 1$ (large
$\tan{\beta}$), where $k_D^2 \sim 1 \pm 0.2$ in Fig.~\ref{fig:crazy3},
while $k_D^2 \sim 1.2 \pm 0.4$ in Fig.~\ref{fig:KD2sin2bma}.  This, as
we shall see, is rooted in the production.

\begin{figure}[h]
\centering
\includegraphics[width=0.60\textwidth]{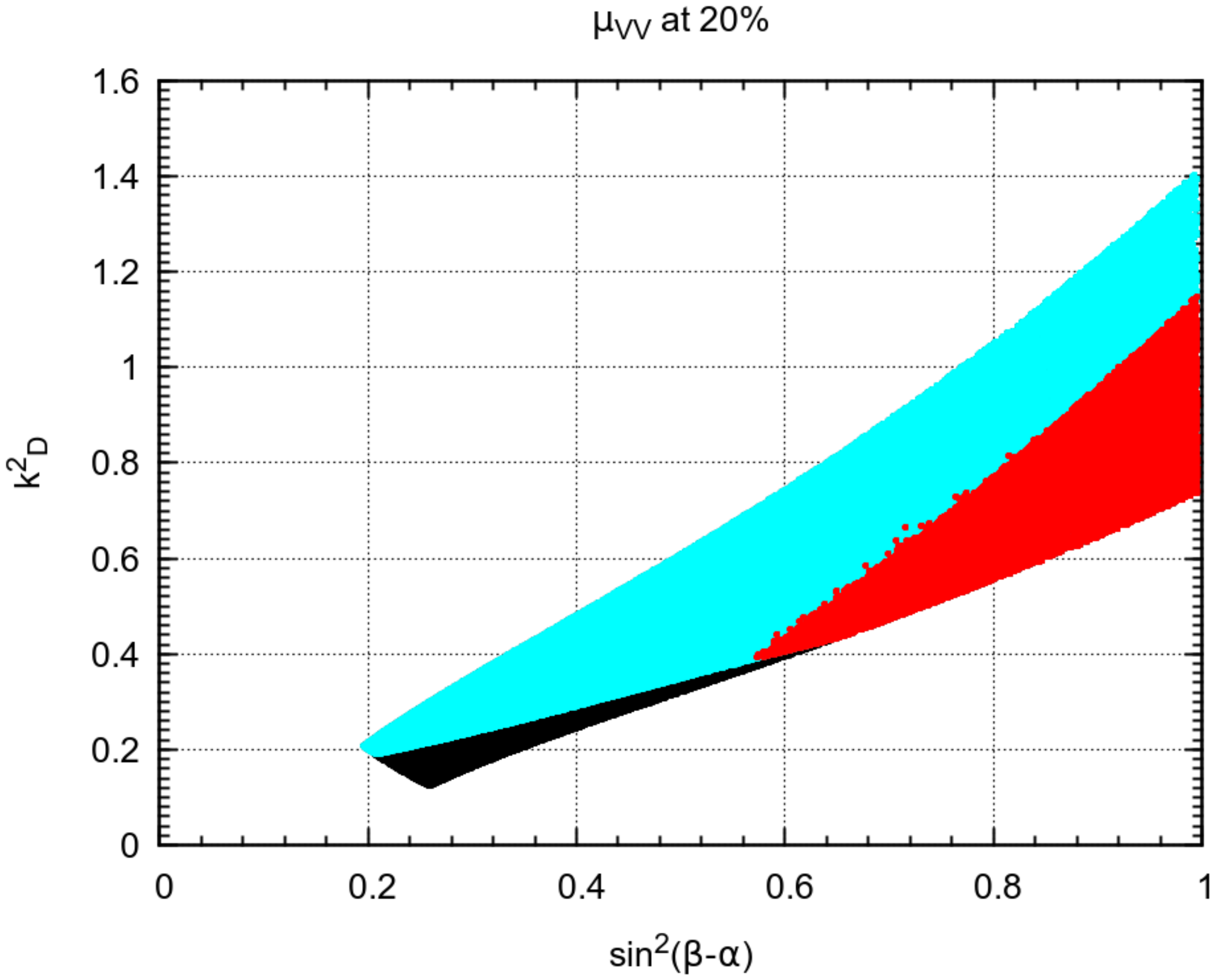}
\caption{Same as in Fig.~\ref{fig:KD2sin2bma}, but for the
  assumed production
  rates in Ref.~\cite{FGHS}. See text for details.}
\label{fig:KD2sin2bma_rui}
\end{figure}

It is also interesting to compare Fig.~\ref{fig:KD2sin2bma}, with
Fig.~\ref{fig:KD2sin2bma_rui}, which we have drawn using the
assumed production rates in Ref.~\cite{FGHS}.
Notice that the values of $k_D^2$ are now
smaller, especially for $\sin^2{(\beta-\alpha)} \sim 1$
(large $\tan{\beta}$).

It is easy to see that imposing further $0.8 \leq \mu_{\tau^+ \tau^-}
\leq 1.2$ may not make a substantial difference.  To understand
qualitatively the impact of channels other than $h \ra VV$, let us
assume that all observed decays will be measured at the SM rates, with
the same error $\delta$.  Using Eqs.~\eqref{mus}-\eqref{RTW}, we find
\be
1 \pm 2 \delta \sim
\frac{\mu_{f_1}}{\mu_{f_2}}
=
\frac{R_D^{f_1}}{R_D^{f_2}}
\label{ratio_mus}
\ee
for all final states $f_1$ and $f_2$.  Notice that this relation does
\textit{not} depend on the production rate, nor on the total width
ratios, which are the same for all decays\footnote{Except $b \bar{b}$,
if we consider that it is only measured in associated production.}.
In particular,
\be
\frac{\mu_{\tau^+ \tau^-}}{\mu_{VV}}
=
\frac{k_D^2}{\sin^2{(\beta-\alpha)}},
\label{ratio_tautau_VV}
\ee
where we have used Eqs.~\eqref{kD}-\eqref{ghVV}.  This means that,
roughly speaking, $k_D^2$ should lie between the lines $k_D^2 = 0.6\,
\sin^2{(\beta-\alpha)}$ and $k_D^2 = 1.4\, \sin^2{(\beta-\alpha)}$,
when we consider points which pass current data at around $20\%$.
Close to $\sin^2{(\beta-\alpha)} \sim 1$, this should reduce the range
of $k_D^2$ from $(0.8,\, 1.6)$ to, roughly $(0.8,\, 1.4)$.  We did the
corresponding simulation
(shown in the cyan/light-gray region of Fig.~\ref{fig:KD2sin2bma})
and find roughly $(0.8,\, 1.3)$.
Notice that adding $h \ra b \bar{b}$, assuming that it is
produced/measured in all channels with the same $20\%$ error, has no
impact, because it would lead to the same Eq.~\eqref{ratio_tautau_VV}.
So, we might as well leave it out.
Before closing the discussion on the $0.8 <\mu_{\tau^+ \tau^-}<1.2$ cut,
let us explain why this has a small effect on
Fig.~\ref{fig:KD2sin2bma},
with our production,
and has almost no effect on Fig.~\ref{fig:KD2sin2bma_rui}.
The reason is that smaller values of $k^2_D$ also
imply that the ratio $\mu_{\tau^+ \tau^-}$ is smaller and this
explains why most points that passed the $\mu_{VV}$ cut at 20\%
(black) also pass the cut in $\mu_{\tau^+ \tau^-}$ (cyan/light-gray).
The black points in Fig.~\ref{fig:KD2sin2bma_rui} are behind
the cyan(light-gray) points and
only appear for small values of $k^2_D$,
due to the lower cut on $\mu_{\tau^+ \tau^-}$.

Fig.~\ref{fig:KD2sin2bma} also shows in red/dark-gray the points generated
from the SET, and constrained by $0.8 \leq \mu_{\gamma\gamma} \leq
1.2$, in addition to the constraints $0.8 \leq \mu_{VV} \leq 1.2$ and
$0.8 \leq \mu_{\tau^+ \tau^-} \leq 1.2$.  Thus, the combination of
$VV$, $\tau^+ \tau^-$, and $\gamma\gamma$ constraints forces
$\sin^2{(\beta-\alpha)} > 0.5 $.  We recall that, from $h \ra VV$
alone, $\sin^2{(\beta-\alpha)} \sim 1 $ for $\tan{\beta} > 10$, with a
minute spread.

We now turn to a qualitative understanding of the impact
of the differing production rates in Figs.~\ref{fig:KD2sin2bma}
and \ref{fig:KD2sin2bma_rui}.
If all production
occurred through gluon fusion with an intermediate top, then the
answer would be that an increase in production rates would have no
impact at all, because it would cancel in Eq.~\eqref{RP}, and we would
still have $R_P = k_U^2$.  In the SM, the production is indeed
dominated by gluon fusion with an intermediate top.  But, for the
gluon fusion in the 2HDM, the interference with an intermediate bottom
becomes important.  Indeed, let us write
\be
\sigma^\textrm{2HDM}(gg \ra h)
=
k_U^2\, g_{tt}
+
k_U\, k_D\, g_{tb}
+
k_D^2\, g_{bb},
\ee
where $g_{bb} \ll |g_{tb}| \ll g_{tt}$.  In the SM, $k_U=k_D=1$, and
$\sigma^\textrm{SM}(gg \ra h) \sim g_{tt}$.  Thus, assuming that all
production goes through gluon fusion, we find from Eq.~\eqref{RP}
\be
R_P \sim
k_U^2
\left[
1 + \frac{k_D}{k_U} \frac{g_{tb}}{g_{tt}}
\right].
\label{RP_approx}
\ee
where we have neglected $g_{bb}$ (we have verified that this is indeed
a very good approximation).  This equation has many features that one
would expect.  If the interference is very small, $k_D g_{tb}/(k_U
g_{tt}) \ll 1$, and we recover $R_P \sim k_U^2$, as mentioned above.
If one were to increase $g_{tb}$ and $g_{tt}$ by the same
multiplicative factor, then $R_P$ would not be altered.
So, what is crucial
in the difference between Figs.~\ref{fig:KD2sin2bma}
and \ref{fig:KD2sin2bma_rui}
is that the mix of $g_{tb}$ and $g_{tt}$ has been
altered between the simulations,
with $|g_{tb}|/g_{tt}$ becoming
larger with the production rates used in this article.
This is more important for large values of
\be
\frac{k_D}{k_U} = - \tan{\alpha} \tan{\beta}
\sim
-\,
\sqrt{\frac{\sin^2{(\beta-\alpha)}}{\mu_{VV}}}.
\label{kD_over_kU}
\ee
The approximation at the end would hold if we were to keep the
assumptions of section~\ref{subsec:VV}.  The first equality in
Eq.~\eqref{kD_over_kU} would lead us to believe that the second term
in Eq.~\eqref{RP_approx} is much more important as $\tan{\beta}$
increases.  However, this is mitigated by the fact that, as the
analysis in section~\ref{subsec:VV} and the approximation at the end
of Eq.~\eqref{kD_over_kU} show, $k_D/k_U$ is tied to
$\sin^2{(\beta-\alpha)}/\mu_{VV}$.  Indeed, the right panel of
Fig.~\ref{fig:5L}, obtained with a full simulation, shows that there
are effects of differing production rates as low as
$\tan{\beta} \sim 1$.
Before proceeding,
it is useful to stress this point.
The intuition gained by looking
at the dependence of the couplings on $\alpha$ and $\beta$,
such as in the first equality in Eq.~\eqref{kD_over_kU},
can be completely altered once some experimental bound is imposed,
such as the $\mu_{VV}$ seen in the approximation at the end
of Eq.~\eqref{kD_over_kU},
because the bound may impose rather nontrivial constraints between
$\alpha$ and $\beta$.
In this case,
for each $\tan{\beta}$,
the range of allowed $\alpha$ is correlated and very small.

Having established that $R_P$ is larger in our simulation than in the
simulation of Fig.~\ref{fig:KD2sin2bma_rui},
we must now understand its differing
impact on $\mu_{\gamma\gamma}$, which is almost the same, and on
$\mu_{b \bar{b}}$, which increases.

The crucial point comes from the previous section, where we found that
$0.8 \leq \mu_{VV} \leq 1.2$ alone gives a very tight constraint on
the possible values of $k_V^2 = \sin^2{(\beta-\alpha)}$, for a given
value of $\tan{\beta}$.  Thus, for fixed $\tan{\beta}$, if we wish to
keep $\mu_{VV} = R_P k_V^2 R_{TW}$ constant and close to one, we must
always keep $R_P R_{TW} \simeq \textrm{constant}$.  As a result, the
only way to accommodate an increased production is to have a decreased
$R_{TW}$ (which is roughly determined by $1/k_D^2$), and to increase
$k_D^2$.  This explains why $k_D^2$ is larger when we use
the larger
$|g_{tb}|/g_{tt}$, as in
Fig.~\ref{fig:KD2sin2bma}, than it is when we use the smaller
production $|g_{tb}|/g_{tt}$, as in
Fig.~\ref{fig:KD2sin2bma_rui}.
Since $k_D^2$ appears in both $h \ra b \bar{b}$ and
$h \ra \tau^+ \tau^-$, both are increased in our
simulation.

If we were to take the right panel of Fig.~\ref{fig:5L}
at face value,
we might have been led to conclude that a measurement
$ 0.9 \leq \mu_{b \bar{b}} \leq 1.1$ or
$ 0.9 \leq \mu_{\tau^+ \tau^-} \leq 1.1$
would already exclude the $k_D < 0$ solution for large
$\tan{\beta}$, as can be seen in the right panel of Fig.~\ref{fig:5L}
(red/dark-gray region).
Unfortunately, as we have shown, these rates are extremely sensitive
to the production
and, thus, cannot be used to exclude $k_D < 0$.

In contrast, because, for fixed $\tan{\beta}$, $\mu_{VV}$ implies
roughly that $R_P R_{TW} \simeq \textrm{constant}$,
$\mu_{\gamma\gamma}$ is virtually independent of the production and only
depends on the decay rate $h\to \gamma\gamma$. As the largest
contribution to this decay comes from the $W$ boson diagrams, and this
coupling is already fixed by $\mu_{VV}$, $\mu_{\gamma\gamma}$ will be
rather insensitive to the QCD corrections in the production and can be used
to constrain $k_D<0$.  As a result, our prediction for
$\mu_{\gamma\gamma}$ in the right panel of Fig.~\ref{fig:5L} mirrors
that
in Fig.~5-Left of Ref.~\cite{FGHS}.

The black points in the right panel of Fig.~\ref{fig:5L}
represent the allowed region for $\mu_{\gamma\gamma}$ when we take
$0.95 \leq \mu_{VV} \leq 1.05$. As the highest value for this range is
only slightly above 0.9,
we agree with the conclusion of
Ref.~\cite{FGHS} that a putative $5\%$ measurement of $h \ra
\gamma\gamma$ at 8 TeV around the SM value would rule out $k_D<0$.

In summary, the constraint $R_P R_{TW} \simeq \textrm{constant}$ means
that, when we increase the $|g_{tb}|/g_{tt}$ mix in the production
rates, $\mu_{\gamma\gamma}$ will stay the
same\footnote{And, indeed, $\mu_{Z \gamma}$.},
as we have found in Fig.~\ref{fig:5L}.
In contrast,
since an increased production implies an increased $k_D^2$,
we find that $\mu_{b \bar{b}} = \mu_{\tau^+ \tau^-} = R_P k_D^2 R_{TW}
\simeq \textrm{constant}\ k_D^2$ must increase, in accordance with
what we see in the same figure.

There are three points to note.
First,
the next LHC run will occur at 14 TeV,
while the current data exists for 8 TeV.
Second, the same argument that
showed that $\mu_{\gamma\gamma}$ is stable against changes in
production can be applied to $\mu_{Z\gamma}$.
Third,
the same delayed decoupling effect found in $\mu_{\gamma\gamma}$
appears in $\mu_{Z\gamma}$.
We address these issues in the next section.

\section{\label{sec:14} Predictions for the 14 TeV run}

Strictly speaking, future LHC experiments will be carried out at 14
TeV.  Moreover, the dominant gluon fusion process shifts by almost a
factor of $2.5$ in going from $8$ to $14$ TeV.  Naively, when
$\tan{\beta}$ becomes large, the interference between the dominant
gluon fusion through a top triangle and the gluon fusion through a
bottom triangle becomes important, and then the sign of $k_D$ is
crucial.
However,
as we have already pointed out, things are complicated
by the fact that $k_D/k_U$ is tied to
$\sin^2{(\beta-\alpha)}/\mu_{VV}$, and current experiments keep
$\sin^2{(\beta-\alpha)}>0.5$.  Moreover, in gluon fusion, the
magnitude squared of the top triangle, the magnitude squared of the
bottom triangle, and the interference are multiplied by almost the
same factor as one goes from 8 to 14 TeV.  As a result, most points
that only differ from the SM model measurements by, say, $20\%$ at 8
TeV will also differ from the SM model measurements by $20\%$ at 14
TeV,
when we use our production based on the current version of HIGLU
with specific PDF's and energy scales.
We have performed a simulation
with 146110 points to test this issue. Only 800 of those (around
$0.6\%$), pass the $20\%$ test at 8 TeV but not at 14 TeV.  So, the
conclusions are unaffected by this issue.

In any case, we perform here the following analysis.  We first find
points (satisfying the conditions in the SET) which differ from the SM
at 8 TeV by $20\%$.  Then, we use those 2HDM points to generate all
rates at 14 TeV.  Our subsequent discussions of the $\mu$ parameters
and, in particular, on the impact of $h \ra Z \gamma$, are only based
on the surviving points.

\begin{figure}[htb]
\centering
\begin{tabular}{cc}
\includegraphics[width=0.49\textwidth]{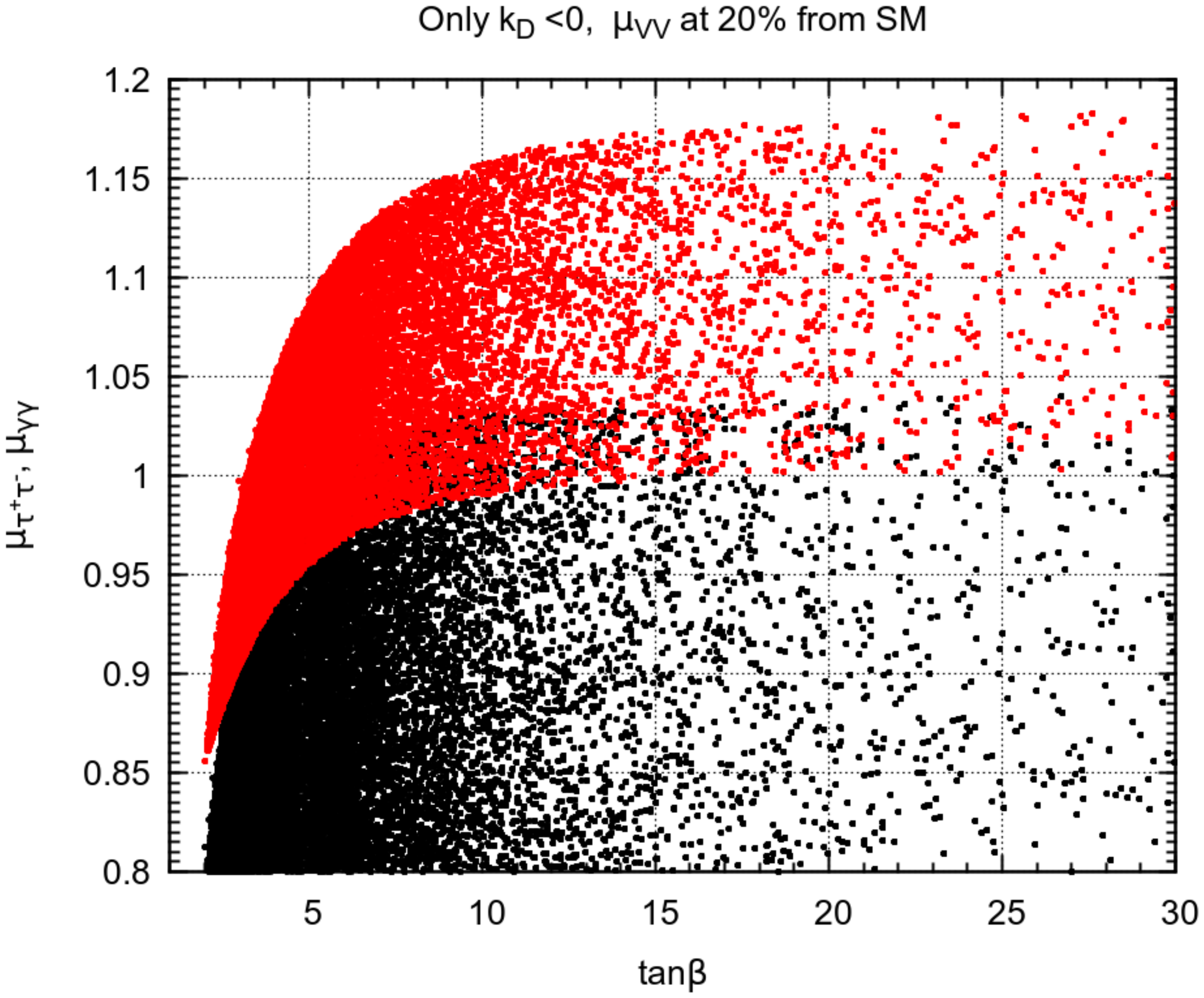} &
\includegraphics[width=0.49\textwidth]{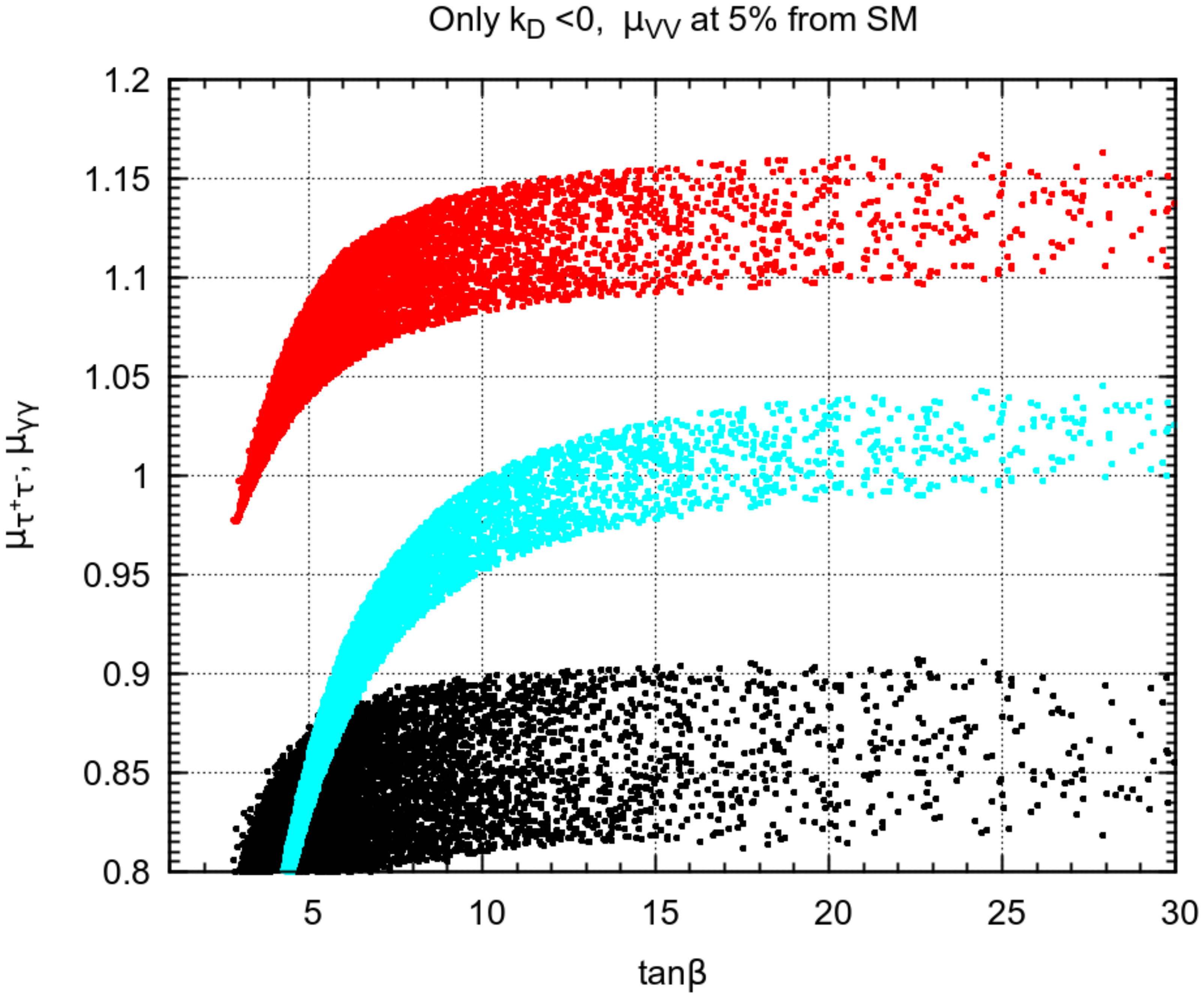}
\end{tabular}
\caption{Left panel: Prediction for $\mu_{\tau^+ \tau^-}$ (red/dark-gray) and
  $\mu_{\gamma\gamma}$ (black) as a function of $\tan\beta$ for the LHC at
  14 TeV with the constraint of $20\%$ errors at 8 TeV.
  Right panel: Assuming now that
  $\mu_{VV}$ are within 5\% of the SM prediction at 14TeV,
  we plot the same quantities.
  Also shown (cyan/light-gray) is the prediction
  for $\mu_{b \bar{b}}(Vh)$ from associated production.}
\label{fig:5L14}
\end{figure}

Assuming current experiments ($20\%$ errors at 8 TeV), our predictions
for $\mu_{\tau^+ \tau^-}$ (in red/dark-gray) and $\mu_{\gamma\gamma}$
(in black) are shown on the left panel of Fig.~\ref{fig:5L14}.
We see that, at this level of precision, we cannot rule out the $k_D<0$
branch.

If we now imagine that, in addition, the $\mu_{VV}$ are measured at 14
TeV to lie around unity with a $5\%$ precision, then we obtain for
$\mu_{\tau^+ \tau^-}$ (in red/dark-gray) and $\mu_{\gamma\gamma}$ (in
black) in the right panel of Fig.~\ref{fig:5L14}.
Here, we would be led to conclude that a $5\%$ measurement of
$\mu_{\tau^+ \tau^-} \sim 1$ would exclude $k_D<0$ for large
$\tan{\beta}$.  As explained in the previous section, this conclusion
is misleading since the $\mu_{\tau^+ \tau^-}$ (and the $\mu_{b
\bar{b}}$ rates, combining all production modes) depend crucially on
the detailed mix of the gluon production through intermediate tops and
bottoms.  Thus, we
agree with Ref.~\cite{FGHS}
that a $5\%$
measurement of $\mu_{\gamma\gamma}$ can be used to exclude the
wrong-sign solution, while $\mu_{\tau^+ \tau^-}$ should not.

We recall that the $\mu_{b \bar{b}}$ we present (in red/dark-gray) in
Fig.~\ref{fig:5L} was calculated
assuming
that $b \bar{b}$ is measured in all channels,
and using our production rates.
In that case, it would seem that a $5\%$ of
$\mu_{b \bar{b}}$ could exclude $k_D <0$.  However, as with
$\mu_{\tau^+ \tau^-}$, the result is very sensitive to the production,
and, thus, cannot be used to probe $k_D<0$.  In foreseeing the 14 TeV
run,
we differ from Ref.~\cite{FGHS}, and study $b \bar{b}$ only in
the $Vh$ production channel,
shown in cyan/light-gray on the right panel of
Fig.~\ref{fig:5L14}.
Unfortunately, in contrast with what
happens with our $\mu_{b \bar{b}}$ in Fig.~\ref{fig:5L}, a $5\%$
measurement of $\mu_{b \bar{b}}(Vh)$ is centered around unity for
$\tan{\beta} > 10$, and, thus, it cannot be used to preclude $k_D<0$.

We now turn our attention to the decay $h \ra Z \gamma$.  As mentioned
above, there are three good reasons to look at this decay.  First, the
decay will be probed at LHC's Run2, and there are already upper bounds
on it from Run1.
Second, as for $\mu_{\gamma\gamma}$, we did not find
a significant difference when using
different production rates.
Third, the
delayed decoupling
that has been used in showing the
usefulness of a future measurement of $\mu_{\gamma\gamma}$ is also
present in $\mu_{Z\gamma}$.  The expressions for this decay can be
found in Ref.~\cite{HHG}, which we have checked.

\begin{figure}[htb]
\centering
\includegraphics[width=0.60\textwidth]{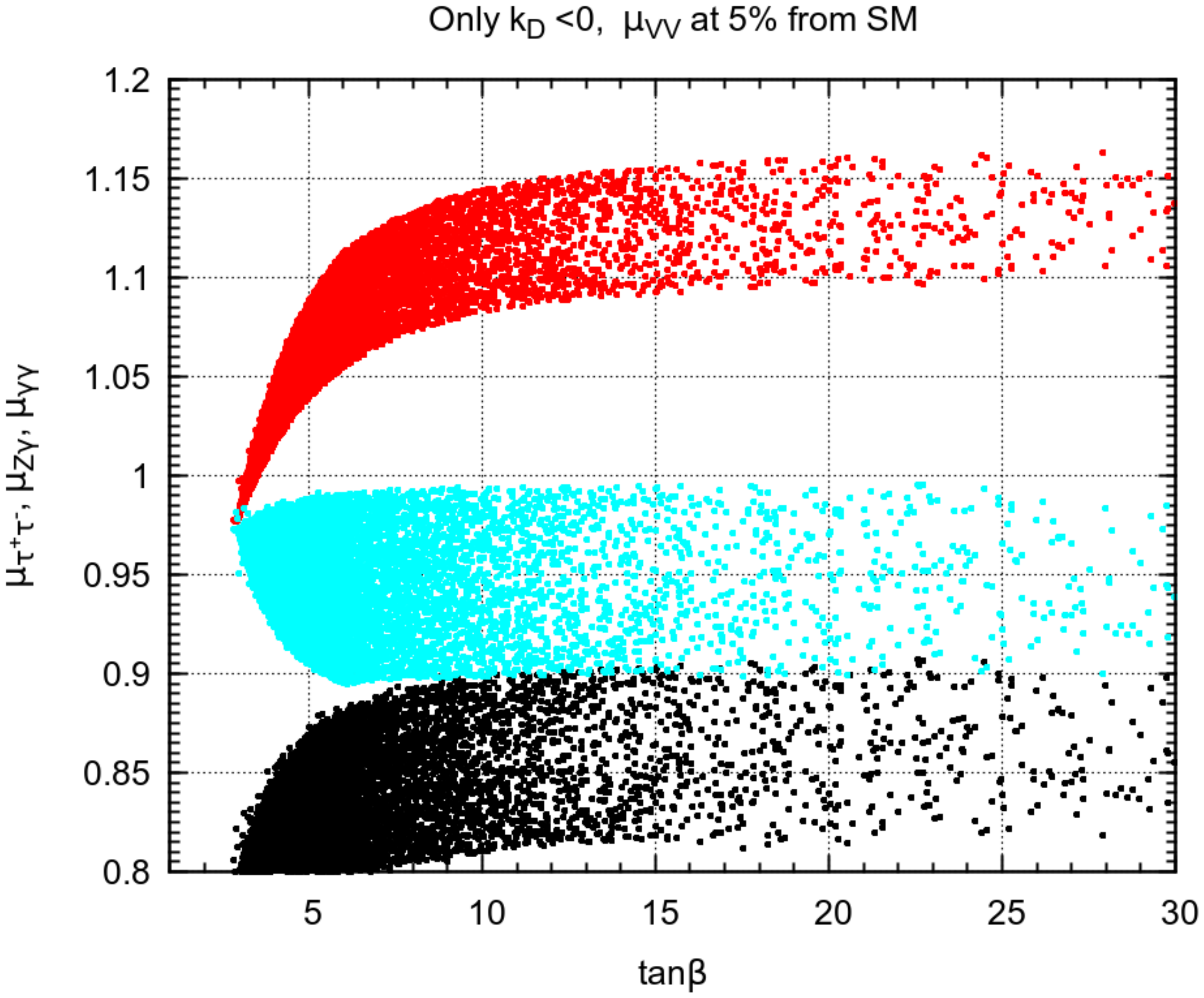}
\caption{Prediction for $\mu_{\tau^+ \tau^-}$ (red/dark-gray),
$\mu_{Z\gamma}$ (cyan/light-gray) and $\mu_{\gamma\gamma}$ (black) as
a function of $\tan\beta$, for the LHC at 14 TeV, with a measurement of
$\mu_{VV}$ within $5\%$ of the SM at 14 TeV.}
\label{fig:Zph}
\end{figure}

Starting from the SET, we calculated $\mu$ for $VV$, $\gamma\gamma$,
and $\tau^+ \tau^-$ at 8 TeV, requiring that all lie within $20\%$ of
the SM.  The remaining points were required to pass $\mu_{VV}$, within
$5\%$ of the SM, at 14 TeV.  We then calculated $\mu_{Z\gamma}$,
$\mu_{\tau^+ \tau^-}$ and $\mu_{\gamma\gamma}$.  Our results are shown
in Fig.~\ref{fig:Zph}.
There are bad news and good news.

\begin{figure}[htb]
\centering
\includegraphics[width=0.60\textwidth]{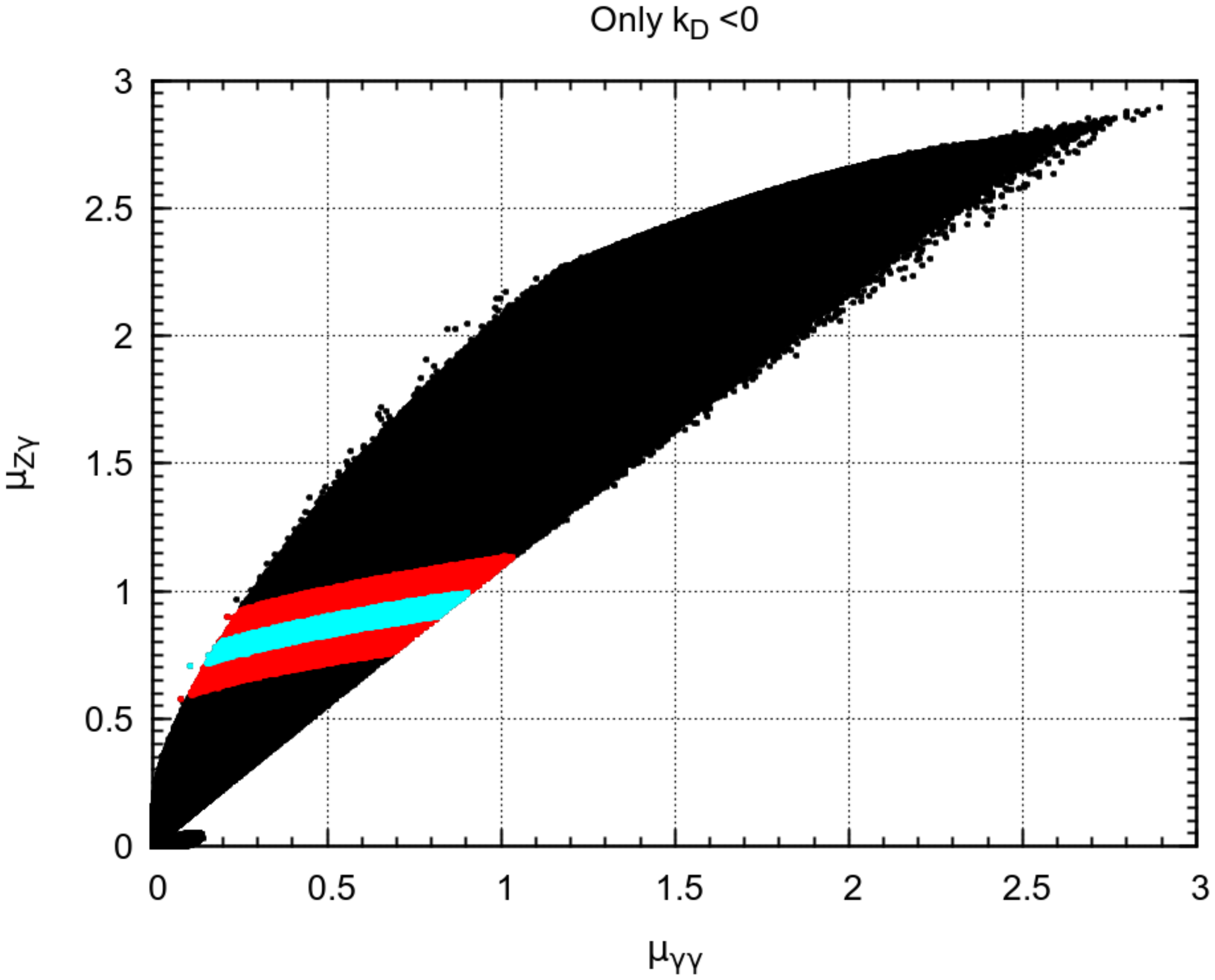}
\caption{Predictions for $\mu_{Z\gamma}$ versus $\mu_{\gamma\gamma}$
at 14 TeV, for $k_D<0$.
In black, we have the points in the SET
(obeying theoretical constraints and $S,T,U$, only).
In red/dark-gray (cyan/light-gray),
the points satisfying in addition $VV$
within $20\%$ ($5\%$)  of the SM, at 14 TeV.}
\label{fig:ellipse_kDneg}
\end{figure}

The bad news comes from the fact that the results in
Fig.~\ref{fig:Zph} show that $\mu_{Z\gamma} \lesssim 1$.
Therefore, this channel cannot be used to exclude the
$k_D<0$ solution.
The good news are the following.
The ratio $\mu_{VV}$, even at 20\%, puts a strong bound
on $\mu_{Z\gamma}$.
In fact, we found that,
for $k_D<0$ and before applying the $\mu_{VV}$ constraint,
$\mu_{Z\gamma}$ could be as large
as two for $\mu_{\gamma\gamma} \sim 1$, as shown in
the black region of Fig.~\ref{fig:ellipse_kDneg}.
However,
the requirement that $\mu_{VV}$ should be within 20\% of the
SM drastically limits this upper bound, requiring it to be very close
to the SM value,
as shown in the red/dark-gray region of Fig.~\ref{fig:ellipse_kDneg}.
If we require a measurement of $\mu_{VV}$ to be within 5\% of the SM
(cyan/light-gray region of Fig.~\ref{fig:ellipse_kDneg}),
then both $\mu_{\gamma\gamma}$  and $\mu_{Z\gamma}$
have to be below their SM values for $k_D<0$.
We find that this effect is more predominant in
$\gamma\gamma$ ($\mu_{\gamma\gamma} < 0.9$)
than in
$Z\gamma$ ($\mu_{Z\gamma} < 1$).

Having discussed what we can learn from $\mu_{\gamma\gamma}$  and
$\mu_{Z\gamma}$ for the wrong-sign branch, $k_D<0$,
we can ask what is the situation with the \textit{normal} branch, $k_D>0$.
\begin{figure}[!htb]
\centering
\includegraphics[width=0.60\textwidth]{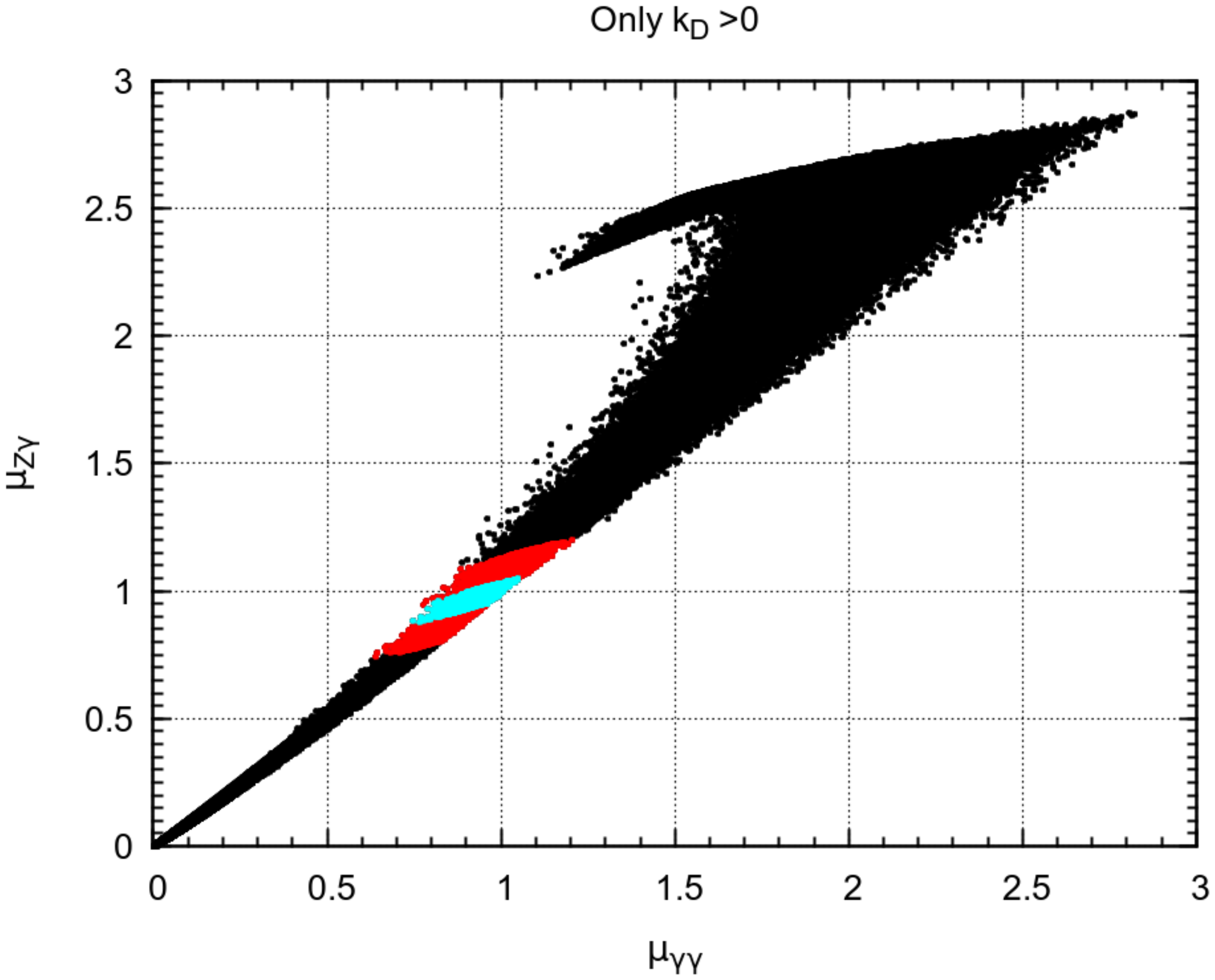}
\caption{Predictions for $\mu_{Z\gamma}$ versus $\mu_{\gamma\gamma}$
at 14 TeV, for $k_D>0$.
In black, we have the points in the SET
(obeying theoretical constraints and $S,T,U$, only).
In red/dark-gray (cyan/light-gray),
the points satisfying in addition $VV$
within $20\%$ ($5\%$)  of the SM, at 14 TeV.}
\label{fig:ellipse_kDpos}
\end{figure}
This is shown in Fig.~\ref{fig:ellipse_kDpos}. We see that even before
requiring any constraint on $\mu_{VV}$ (black points),
there is only a very small region with
large  $\mu_{Z\gamma}$ which is compatible with
$0.8 \leq \mu_{\gamma\gamma} \leq 1.2$ from current LHC data.
In particular,
points from the SET, with  $\mu_{\gamma\gamma} \sim 1$
and $\mu_{Z\gamma} \sim 2$,
allowed for $k_D<0$ in the black region of
Fig.~\ref{fig:ellipse_kDneg},
are almost forbidden for $k_D>0$ in the black region of
Fig.~\ref{fig:ellipse_kDpos}.
If we further require $\mu_{VV}$ to be
within 20\% (red/dark-gray) or 5\% (cyan/light-gray) both $\mu_{\gamma\gamma}$  and
$\mu_{Z\gamma}$ have to be close to the SM values,
with a wider range allowed for $\mu_{\gamma\gamma}$.

We conclude that,
for both signs of $k_D$,
current bounds on $\mu_{VV}$ already preclude a
value of $\mu_{Z\gamma} > 1.5$ from being compatible with
the usual 2HDM with softly broken $Z_2$.
A measurement in the next LHC run of $\mu_{VV}$ lying
within $5\%$ of the SM will essentially force $\mu_{Z\gamma} \lesssim 1$ for
$k_D < 0$ and $\mu_{Z\gamma} \lesssim 1.05$ for
$k_D > 0$.

\section{\label{sec:flipped}Predictions for the Flipped 2HDM}

In this section,
we analyze the Flipped 2HDM.
This coincides with the Type II 2HDM, except
that the charged leptons couple to the Higgs
proportionally to $k_U$ (not $k_D$).

\begin{figure}[htb]
\centering
\begin{tabular}{cc}
\includegraphics[width=0.49\textwidth]{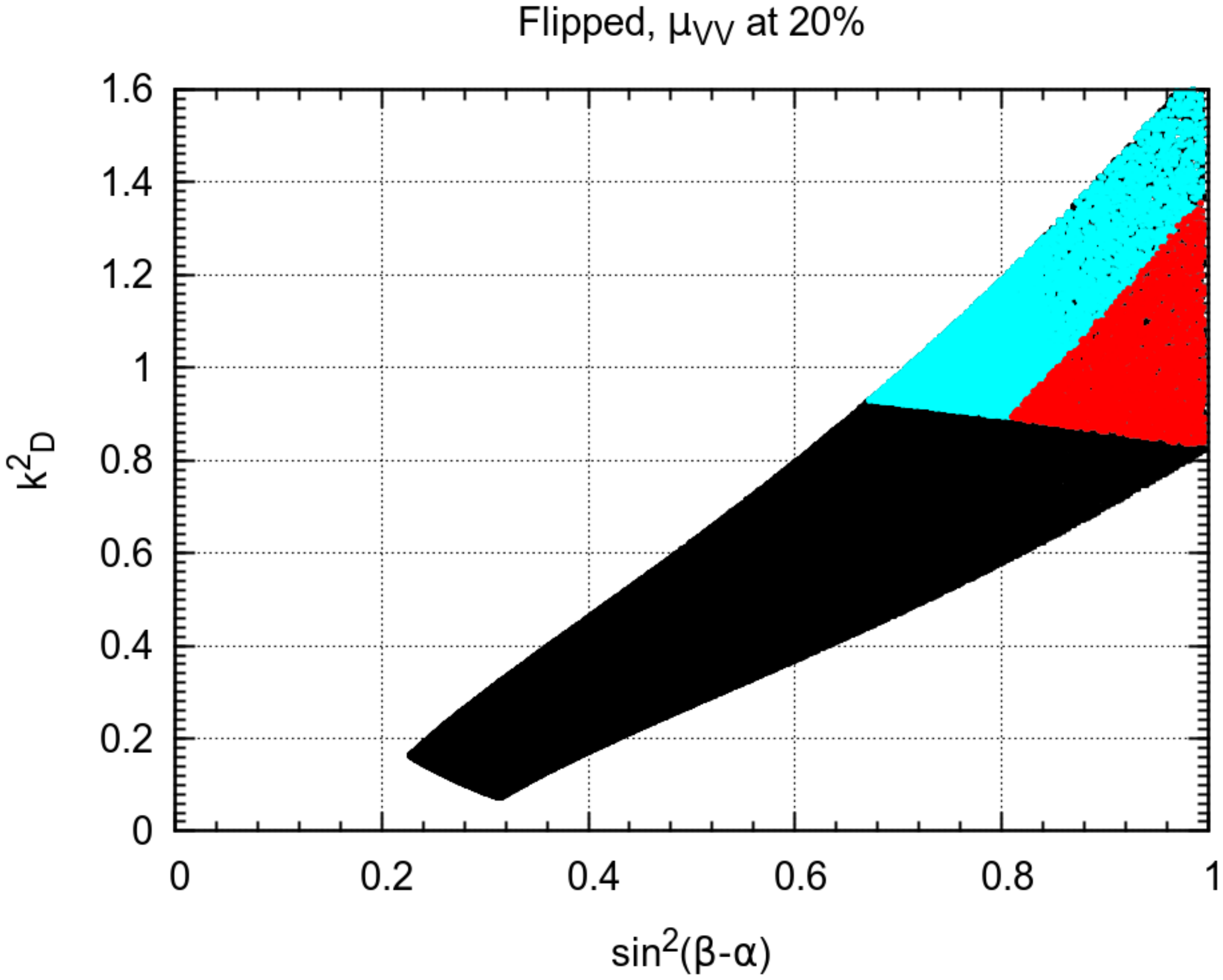} &
\includegraphics[width=0.49\textwidth]{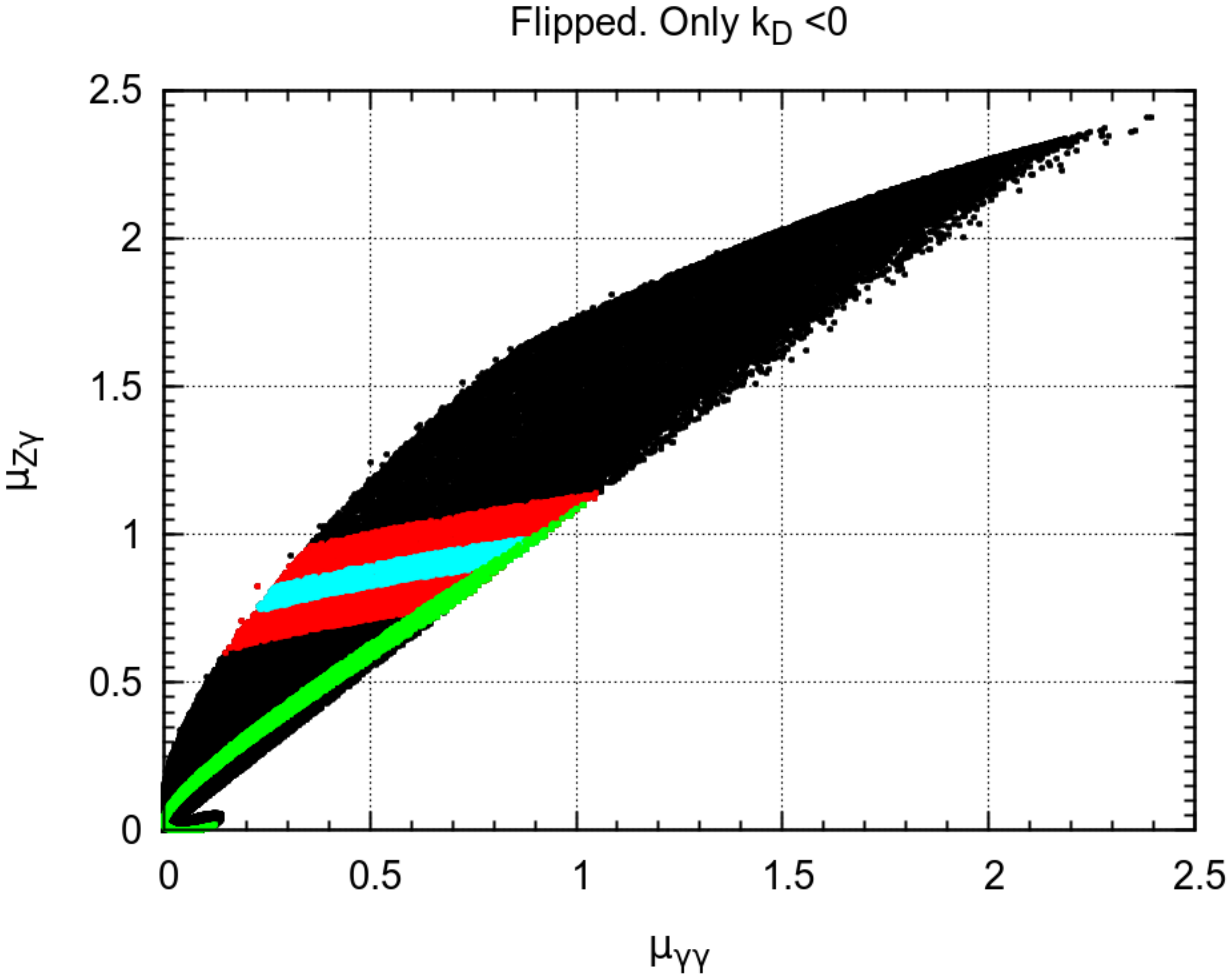}
\end{tabular}
\caption{Left panel: Allowed region for $k^2_D$ as a function of  $\sin^2{(\beta-\alpha)}$
in the Flipped 2HDM, for all
  points with $k_D<0$ that obey $0.8 \leq  \mu_{VV} \leq 1.2$
  (black). The region in cyan (light-gray)
  is obtained by imposing in addition that
  $0.8 \leq  \mu_{\tau\tau} \leq 1.2$, while in the region in red (dark-gray) we
  further impose  $0.8 \leq  \mu_{\gamma\gamma} \leq 1.2$.
  Right panel: Predictions for $\mu_{Z\gamma}$ versus $\mu_{\gamma\gamma}$
  at 14 TeV, for $k_D<0$,
  in the Flipped 2HDM.
  In black, we have the points in the SET
  (obeying theoretical constraints and $S,T,U$, only).
  In red/dark-gray (cyan/light-gray),
  the points satisfying in addition $VV$
  within $20\%$ ($5\%$)  of the SM, at 14 TeV.
  Shown in green/light-gray are the points satisfying
  $\mu_{\tau^+ \tau^-}$ at $20\%$ of the SM,
  which lie on a line going diagonally from the
  origin with almost unit slope.}
\label{fig:Flipped}
\end{figure}

We recall that $\mu_{\tau^+ \tau^-}$ does not
have a big effect in Fig.~\ref{fig:KD2sin2bma},
for the Type II 2HDM.
This has a simple explanation,
through the approximation in Eq.~\eqref{ratio_tautau_VV}.
In the Flipped 2HDM,
the same approximation yields
\be
\frac{\mu_{\tau^+ \tau^-}}{\mu_{VV}}
=
\frac{k_U^2}{\sin^2{(\beta-\alpha)}},
\label{ratio_tautau_VV_flipped}
\ee
leading one to suspect that $\mu_{\tau^+ \tau^-}$
might have a larger effect here.
This is confirmed in the left panel of
Fig.~\ref{fig:Flipped},
where we show our 8 TeV results for $k_D^2$ as
a function of $\sin^2{(\beta-\alpha)}$.
The colour codes,
explained in the figure caption, mirror those in
Fig.~\ref{fig:KD2sin2bma}.
Here the $20\%$ measurement of $\mu_{\tau^+ \tau^-}$
does have a big impact.

However,
one might suspect that this may not change much the conclusions
on $\gamma\gamma$ and $Z \gamma$,
because,
as mentioned before,
those were primarily determined by the constraint on
$\mu_{VV}$.
This is what we find in the right panel of
Fig.~\ref{fig:Flipped}.
The effect of $\mu_{\tau^+ \tau^-}$ is to
reduce the allowed region by a very fine slice,
shown in the right panel of Fig.~\ref{fig:Flipped} as
a green/light-gray line going diagonally from the
origin with almost unit slope.
This figure should be compared with Fig.~\ref{fig:ellipse_kDneg},
which holds in the Type II 2HDM.
In both cases,
a $5\%$ measurement of $\mu_{\gamma\gamma}$
($\mu_{Z\gamma}$)
will (will not) exclude $k_D<0$.

\section{\label{sec:conclusions}Conclusions}

We have analysed the Type II 2HDM with softly broken $Z_2$,
scrutinizing the possibility that the $hb\bar{b}$ coupling has a sign
opposite to that in the SM and the impact on this issue of $Z\gamma$.
We impose the usual theoretical constraints,
assuming that $\mu_{VV}$, $\mu_{\tau^+ \tau^-}$,
and $\mu_{\gamma\gamma}$ differ from the SM by no more than $20\%$
at 8 TeV.
We found that the constraint from $\mu_{VV}$ is crucial,
and can be understood in simple trigonometric terms.
In particular,
we show that this cut has a rather counter-intuitive implication.
Before this cut is applied,
it would seem that the importance of the bottom-mediated
gluon fusion production mechanism would grow linearly with $\tan{\beta}$.
However,
after current bounds are placed on $\mu_{VV}$,
the importance of the bottom-mediated
gluon fusion production mechanism grows asymptotically
into a constant, for large $\tan{\beta}$.
This generalizes as a cautionary tale:
applying a new experimental bound may force unexpected
relations among the parameters,
and the theoretical intuition must be revised
in this new framework.

In projecting to the future,
we have then simulated our points at 14 TeV,
highlighting the fact,
for the issues that interest us,
using the current version of HIGLU at 14 TeV or at
8 TeV leads to the same results.
We have shown that results for the $b \bar{b}$ and
$\tau^+ \tau^-$ depend sensitively on the ratio $g_{tb}/g_{tt}$
encoding the relative weight of the square of the top-mediated
gluon fusion production amplitude, and the interference of
this amplitude with the bottom-mediated
gluon fusion production amplitude.
As a result, these channels should not be used to probe
the $k_D<0$ possibility.
Even if that were not the case,
since $b \bar{b}$ is only measured in associated production
and, as we show, $\mu_{b \bar{b}} (Vh)$ includes unity,
this channel would not be useful.

In contrast,
in our simulations both $\gamma\gamma$ and $Z\gamma$ are roughly
independent of $g_{tb}/g_{tt}$.
In addition,
they exhibit delayed decoupling in the $h H^+ H^-$ vertex.
As a result, they could, in principle, be used to probe
the $k_D<0$ possibility.
Indeed,
as found in Ref.~\cite{FGHS},
a 5\% measurement of $\mu_{\gamma\gamma}$ around unity
will be able to exclude $k_D<0$.

We then performed a detailed analysis of $Z\gamma$.
We show that,
before including the LHC data,
values of $\mu_{\gamma\gamma}$
and $\mu_{Z\gamma}$ were allowed between $0$ and $3$,
but with a correlation between the two,
as shown in the black regions of Fig.~\ref{fig:ellipse_kDneg}
and Fig.~\ref{fig:ellipse_kDpos}.
This correlation is more important
(that is, the region in the figure is smaller)
for $k_D>0$ than it is for $k_D<0$.
In particular,
$\mu_{\gamma\gamma} \sim 1$ with $\mu_{Z\gamma} \sim 2$
would be possible in the latter case,
but not in the former.
Things change dramatically when the simple constraint
$0.8 < \mu_{VV} < 1.2$ is imposed.
In that case,
we obtain the red/dark-gray regions of
Fig.~\ref{fig:ellipse_kDneg} ($k_D<0$)
and Fig.~\ref{fig:ellipse_kDpos} ($k_D>0$).
This already places
$\mu_{\gamma\gamma}$ and $\mu_{Z\gamma}$
close to the SM,
although,
strictly speaking,
points with $\mu_{\gamma\gamma} = 1$ with $\mu_{Z\gamma} = 1$
are not allowed in our simulation when $k_D<0$.
A $5\%$ measurement of $VV$ around the SM at 14 TeV
will bring $\mu_{Z\gamma}$ closer to unity,
for $k_D>0$,
and just below unity,
for $k_D<0$.
Thus,
this decay cannot be used to exclude $k_D<0$.

But we have the reverse advantage.
It is obvious that a measurement of $\mu_{Z\gamma} > 1$ would
exclude the SM.
We have shown that a 5\% precision on $\mu_{VV}$ around the SM,
together with $\mu_{Z\gamma} > 1$,
would also exclude $k_D<0$,
and, together with $\mu_{Z\gamma} > 1.1$,
would exclude altogether the Type II 2HDM with softly broken $Z_2$.
If $\mu_{Z\gamma}$ turns out to lie a mere $20\%$ above the SM value,
then the softly broken Type II 2HDM is not the solution.

Finally,
we analyzed the Flipped 2HDM.
Although there is a substantial difference in the
$k_D^2$ versus $\sin^2{(\beta-\alpha)}$ plane,
this does not change dramatically the $\mu_{\gamma\gamma}$--$\mu_{Z\gamma}$
correlation.
As a result,
here $5\%$ measurements of $VV$ and $\gamma\gamma$
around the SM at 14 TeV will be enough to exclude $k_D<0$,
while $\mu_{Z\gamma}$ will not.

\begin{acknowledgments}

We are grateful to Rui Santos for many discussions related to the
Higgs production channels and to the work \cite{FGHS}.  This work was
partially supported by FCT - \textit{Funda\c{c}\~{a}o para a
Ci\^{e}ncia e a Tecnologia}, under the projects
PEst-OE/FIS/UI0777/2013 and CERN/FP/123580/2011.  D.~F. is also
supported by FCT under the project EXPL/FIS-NUC/0460/2013.

\end{acknowledgments}


\end{document}